\documentstyle[aps,prl,epsf,floats,multicol,amssymb,tighten]{revtex}

\begin{document}

\newcommand{\fig}[2]{\epsfxsize=#1\epsfbox{#2}} \reversemarginpar 
\newcommand{\mnote}[1]{$^*$\marginpar{$^*$ {\footnotesize #1}}}

\def\oppropto{\mathop{\propto}} 
\def\opsimeq{\mathop{\simeq}}
\def\opoverderline{\mathop{\overline}}
\def\operarrow{\mathop{\longrightarrow}}
\def\opsim{\mathop{\sim}}

\bibliographystyle{prsty}

\title{Exact solutions for the statistics of extrema of some random 1D landscapes, Application to the equilibrium and the dynamics of the toy model.} 

\author{Pierre Le Doussal}
\address{CNRS-Laboratoire de Physique Th\'eorique de l'Ecole\\
Normale Sup\'erieure, 24 rue Lhomond, F-75231
Paris}

\author{C\'ecile Monthus}
\address{Service de Physique Th\'eorique,
CEA Saclay, 91191 Gif-sur-Yvette, France}

\maketitle
\begin{abstract}
The real-space renormalization group (RSRG) method
 introduced previously for the Brownian landscape 
is generalized to obtain the joint probability distribution of
 the subset of the important extrema at large scales
of other one-dimensional landscapes.
For a large class of models we give exact solutions
obtained either by the use of constrained path-integrals in the continuum limit, or by solving the RSRG equations via an Ansatz
which leads to the Liouville equation.
We apply in particular our results to the toy model energy landscape, 
which consists in a quadratic potential plus a Brownian potential,
which describes, among others, the energy of a single domain 
wall in a 1D
random field Ising model (RFIM) in the presence of a field gradient.
The measure of the renormalized landscape 
is obtained  explicitly in terms of Airy functions, and
allows to study in details the Boltzmann equilibrium of a particle 
at low temperature as well as its non-equilibrium dynamics. 
For the equilibrium, we give results for the statistics
of the absolute minimum which dominates at zero temperature,
and for the configurations with nearly degenerate minima which govern
the thermal fluctuations at very low-temperature.
For the dynamics, we compute the distribution over samples
 of the equilibration time,
or equivalently the distribution of the largest barrier in the system. We also
study the properties of the rare configurations presenting an
anomalously large equilibration time which govern the
long-time dynamics. We compute the disorder averaged
diffusion front, which interpolates between 
the Kesten distribution of the Sinai model at short rescaled time
and the reaching of equilibrium at long rescaled time. Finally,
the method allows to describe the full coarsening (i.e. many domain walls) of the 1D RFIM in a field gradient as well as its equilibrium. 

\end{abstract}


\widetext

\newpage



\section{Introduction}

\subsection{Overview}

Disordered systems \cite{droplet,replica} 
are usually characterized by a complicated 
energy (or free energy) landscape, with many local minima
representing metastable states, separated by saddles and barriers which
govern the dynamics. In general,  
obtaining analytical information on the statistics of large scale extrema
 and saddles of a random energy landscape $U(x)$ is an extremely hard task.
 In one dimension however, the situation is simpler since one
 has to deal with an alternating
sequence of minima and maxima, and there is hope for analytical progress.
Recently, the large scale extrema of
the pure Brownian landscape and of the biased
Brownian landscape
have been studied with a powerful real space
renormalization technique \cite{us_prl,us_long}.
There the interest was in the non-equilibrium Arrhenius dynamics of a 
particle in such a landscape, the so-called Sinai 
model \cite{sinai,kesten,derrida_pomeau,ledou_1d}. 
At large time this dynamics is dominated by
jumps between deep minima. The renormalization procedure
 thus consists in erasing iteratively 
the small barriers in the landscape, and retaining 
only the large scale, large energy features
which control the large time dynamics. 
This procedure yields the asymptotically exact large time
dynamics in the case of the Sinai model,
 and allows to obtain a host of exact results 
\cite{us_prl,us_long}.
This real space renormalization
 technique has also been used in other one dimensional models,
such as quantum spin chains \cite{ma_dasgupta,dsfrg}, 
reaction diffusion with disorder \cite{us_rd} and 
the random field Ising model \cite{us_long_rf}, with the same fixed points
associated to the pure or to the biased Brownian landscape.
For the coarsening of pure 1D $\Phi^4$ model 
at zero temperature from a random initial condition \cite{derridaphi4},
another fixed point was found which corresponds 
to the successive elimination of smallest domains, the only variable there being the length.
A question is then whether other energy landscapes can be studied using similar techniques.

\subsection{Definition and properties of the toy model}

A one-dimensional energy landscape of interest is 
the so-called ``toy model" defined by the Hamiltonian
\begin{eqnarray}
U_{toy}(x)=\frac{\mu}{2} x^2 + V(x)
\label{deftoy}
\end{eqnarray}
where the random potential $V(x)$ is a Brownian motion with
correlations
\begin{eqnarray}
\overline{ \left(V(x)-V(y) \right)^2 }= 2  \vert x-y \vert
\label{corre}
\end{eqnarray}
 Here we have defined the model directly in the continuum limit, 
but the universality class contains all models with a small scale 
cutoff and the above behavior (\ref{corre}) at large scale.
This model has been introduced by Villain {\it et al.}  
as a zero-dimensional toy-model for an interface
in the random field Ising model \cite{villain83}.
Let us briefly review previous studies of this model and
explain some motivations to reexamine this model.

\subsubsection{Equilibrium : Imry-Ma argument and `statistical tilt symmetry'}

For a single particle,
the Boltzmann equilibrium problem at temperature $T$
is defined by the partition function
$Z= \int dx e^{- U_{toy}(x)/T}$.
Although it seems very simple at first sight, it is an interesting example
of disordered system. Because of the quadratic well,
 the typical position of the particle 
is finite, but it fluctuates in a non trivial way from sample to sample. 
An Imry-Ma argument gives the following scaling
for the typical scale of the absolute minimum
\begin{eqnarray} 
x \sim \left( \overline{< x^2 >_{T=0}} \right)^{1/2} = C \mu^{-2/3} 
\label{scaling1}
\end{eqnarray}
where we denote by the brackets $<.>$ the thermal Boltzmann measure with potential (\ref{deftoy})
and by an overbar the average over the random potential (\ref{corre}).
 The amplitude $C$
will be computed here. Usual methods such as perturbation expansion
(even summing all terms) \cite{villain88} or iteration method \cite{villainsemeria}
completely fail to recover (\ref{scaling1}).
 A Larkin type perturbation \cite{larkin70} with a small scale 
cutoff also predicts $x \sim (1/\mu)$ instead of (\ref{scaling1}).
Within replica variational {\it approximations} of the present model 
(or within
the solvable mean field limit of large embedding space dimension $N \to \infty$, see below) 
this is cured at the price of spontaneous replica symmetry breaking (RSB)
\cite{mezardparisi92,engel}. RSB is of course not expected to occur
in the present exact model (\ref{deftoy} - \ref{corre}) (see Ref. \cite{carpentier} for discussion of the cases where
RSB may occur in a one dimensional model). 

The toy model also possesses the `statistical tilt symmetry' 
as other random field models 
\cite{identities88} or the directed polymer model \cite{huse,hwa},
which imply exact remarquable identities \cite{identities88} between
the disorder averages of the thermal cumulants of the position
at any temperature, as a consequence from the 
statistical translation invariance of the random potential.
These exact identities can be summarized by \cite{identities88}
\begin{eqnarray}
 \overline { \ln < e^{- \lambda x } > } && = T \frac{\lambda^2}{2  \mu}
 \label{genlambda2}
\end{eqnarray}
which is simply quadratic in $\lambda$, so that 
the second cumulant reads
\begin{eqnarray}
&& \overline { < x^2 > - <x>^2 } = \frac{T}{\mu } \label{second}
\end{eqnarray}
and the disorder-average of all thermal cumulants 
of order greater than two vanish!
As emphasized in \cite{identities88},
in the low temperature regime, the result (\ref{second})
implies that the thermal fluctuations of the toy model
are related to the presence of metastable states 
in rare disordered samples.

\subsubsection{Non-equilibrium dynamics : between Sinai and approach of equilibrium}

An Arrhenius dynamics can be defined for the toy model,
 and provided $\mu$ is small so that barriers can be large, 
one can ask about universal long time properties. This
was studied for the special case $\mu=0$ where one recovers the Sinai model \cite{us_prl,us_long}.
The dynamics for small but finite $\mu$ is interesting
 for several physical systems of
the random field type, and has not been studied up to now, to our knowledge. 
Physically, one expects the diffusion to interpolate
between a Sinai-type diffusion at short rescaled time
and the approach of equilibrium at long rescaled time.

\subsection{ Applications of the toy model}

\subsubsection{Directed polymers in random media}

The toy model has also attracted a lot of attention in
connection with the problem of $D$-dimensional manifolds in 
 $(N+D)$ random media, where it plays the role of a prototype model
since it represents the simplest case $D=0$ and $N=1$
\cite{mezardparisi91,mezardparisi92}. 
It has been moreover argued in \cite{parisicorresdp,bocorresdp}
that the toy model actually describes the statistics of
the end-point of a random polymer of length $L$
in $(1+1)$ with delta correlated noise in the limit $L$ large
with the correspondence $\mu = 1/L$. It was found 
numerically in \cite{mezardcorresdp}
that the correlation of the effective random potential for the
directed polymer model behaves as (\ref{corre}) only at short
distances but saturates at large distances. This analysis has been
confirmed by a scaling argument in \cite{huse} and was
finally rigorously proven in \cite{spohn} where the full
statistics of the effective potential was exactly computed
to be an ``Airy process", defined there. Thus the approximation of 
end-point of the directed
polymer in $(1+1)$ by the toy model is valid only locally.

\subsubsection{Burgers turbulence}

The toy model is related to the statistical properties
of a velocity field $v(x,t)$ evolving according to the 
Burgers equation \cite{burgers,frachebourgmartin}
\begin{eqnarray}
 \partial_t v(x,t) +v(x,t) \partial_x v(x,t) = \nu \partial_x^2  v(x,t)
\label{burgers}
\end{eqnarray}
This equation can be solved via the Hopf-Cole transformation as 
\begin{eqnarray}
 v(x,t) = \frac{ \int_{-\infty}^{+\infty} dy \left( \frac{x-y}{t} \right)
e^{- \frac{1}{2 \nu} \left(  \frac{(x-y)^2}{2 t} +\phi(y)  \right)} }
{ \int_{-\infty}^{+\infty} dy 
e^{- \frac{1}{2 \nu} \left(  \frac{(x-y)^2}{2 t} +\phi(y)  \right)} }
\end{eqnarray}
where $\phi(y)= \int_0^y dy' v(y',0)$ is a Brownian motion
in the case where the initial velocity field $v(y',t=0)$
is a white noise process.
In the inviscid limit $\nu \to 0$, the integrals are dominated by 
the saddle   
\begin{eqnarray}
 \xi(x,t) = {\rm min}_y \left(  \frac{(x-y)^2}{2 t} +\phi(y)  \right)
\label{eqsaddle}
\end{eqnarray}
which gives
\begin{eqnarray}
 v(x,t) = \frac{x-\xi(x,t)}{t}
\label{vsaddle}
\end{eqnarray}
The velocity distribution and the statistical properties of shocks
(which appear at the points $x$ where
 there are two degenerate minima in (\ref{eqsaddle})) 
are thus directly related to the properties of minima of the toy model 
(\ref{deftoy}) with the correspondence $\mu=1/t$. 

\subsubsection{Deterministic KPZ equation }

The toy model is also related to the relaxation of an interface
evolving according to the deterministic KPZ equation \cite{kpz,esipov-newman}
\begin{eqnarray}
\partial_t h(x,t) = \nu \partial^2_x h + \frac{\lambda}{2}
(\partial_x h(x,t) )^2 
\label{kpz}
\end{eqnarray}
from a random initial condition $h(y,t=0)$ which is a white noise in $y$.
Indeed, the mapping \cite{kpz,esipov-newman}
\begin{eqnarray}
v(x,t)=- \lambda \partial_x h(x,t)
\end{eqnarray}
transforms the deterministic KPZ equation (\ref{kpz}) into
the burgers equation (\ref{burgers}), leading to
\begin{eqnarray}
 h(x,t) = \frac{2 \nu}{ \lambda}
\ln \int_{-\infty}^{+\infty} \frac{dy}{\sqrt{4 \pi \nu t}}
e^{- \frac{1}{2 \nu} \left(\frac{(x-y)^2}{2 t} + \lambda h(y,0) \right) } 
 \end{eqnarray}
Again in the limit $\nu \to 0$ where the saddle method
 yields (\ref{vsaddle}), the profile $h(x,t)$ is made of parabolic segments
joined together by cusps corresponding to the shocks of the Burgers equation
(see for instance Fig.1 in \cite{kpz}).

\subsubsection{Coarsening and equilibrium in some random field Ising model}

For definiteness, we will consider in the following an Ising spin chain,
but one can study not only spin systems, but also 
for instance liquid helium meniscus on a rough substrate \cite{helium}, 
which does also see random field disorder, and can in principle be probed in confined, quasi one dimensional geometry. 

Let us consider as in \cite{us_long_rf} the 1D random field Ising system of
Hamiltonian
\begin{eqnarray} 
H = - J \sum_{k=1}^{N-1} S_k S_{k+1} - \sum_{k=1}^{N} h_k S_k
\end{eqnarray}
with spins $S_k =\pm 1$, and study its equilibrium and its low temperature Glauber dynamics.
In the domain wall representation, two types of domains walls 
$A$, $(+|-)$, and $B$, $(-|+)$,
in numbers $N_A$ and $N_B$, alternate at
positions $a_\alpha$, $b_\alpha$, respectively, and the Hamiltonian
can be rewritten as
\begin{eqnarray} 
H = H_{ref} + 2 J (N_A + N_B) - \sum_{\alpha=1}^{N_A} U(a_\alpha) + \sum_{\alpha=1}^{N_B} U(b_\alpha)
\end{eqnarray}
with $|N_A-N_B|<1$. The potential felt by domains walls of type B is
\begin{eqnarray} 
U(k) = 2 \sum_{j=1}^k h_j
\end{eqnarray}
while domain walls $A$ feel potential $- U(k)$. 
The dynamics proceeds by diffusion and annihilation of domains of opposite types.

In \cite{us_long_rf}, the zero field $\overline{h_k}=0$
and the uniform applied field $\overline{h_k}=H$ situations
have been studied. The toy model corresponds to the 
other interesting case
\begin{eqnarray}
&& \overline{h_k} = k h  \\
&& \overline{h_k^2} - \overline{h_k}^2 = g
\end{eqnarray}
representing {\it a field gradient} with zero net applied field
(such gradients are easy to produce experimentally).
Indeed, it is easy to see that defining the scaled length $x=2 g k$, 
each domain wall $B$ sees the toy model energy landscape $U(k)=U_{toy}(x)$ with
$U_{toy}(x)$ defined in (\ref{deftoy}) in terms of a unit Brownian and 
\begin{eqnarray}
&& \mu = \frac{ h}{2 g^2} 
\end{eqnarray}
For a positive field gradient ($h>0$), the domain walls of type $B$
$(-|+)$ will be attracted towards the minima of $U_{toy}(x)$, 
and will thus be confined in the neighborhood of 
the point $x=0$ where the applied field vanishes,
whereas the domain walls
of type $A$ $(+|-)$ will be attracted by the local maxima 
of $U_{toy}(x)$ or will be expelled towards the boundaries.

The equilibrium of the toy model considered above
 and its dynamics involving only one particle corresponds to 
a {\it single} domain wall $B$. If the field gradient is $h>0$ and
if the system is initially prepared with the ``natural" boundary conditions, namely
spin far on the left are $(-)$, far on the right $(+)$ and with a {\it single}
domain wall of type $B$ in between, it remains such, until equilibration time $T \ln t_{eq0} = 2 J$ (or
$T \ln t_{eq0} = J$ if domains can be created at boundaries).
During that window of time the domain wall diffuses
 in the toy model landscape and
reaches thermal equilibrium at time $t_{eq}$, a time scale which can be much smaller than 
$t_{eq0}$ if $J$ is sufficiently large. 
Another interesting situation corresponds to {\it unstable} single domain wall dynamics 
and is obtained if the field gradient is applied
against these boundary conditions, e.g. a single $A$ domain initially. The domain is
then expelled towards either side after a time $t_{exp}$ which again can be 
well before $t_{eq0}$ if $J$ is sufficiently large.

Finally, the full, many domain coarsening can be studied and is expected to exhibit 
a large degree of universality in the small $h$, large Imry Ma length $4 J^2/g$, small temperature
$T \ll J$, large time limit as in \cite{us_long_rf}. It is interesting to study because of the competition between the
disorder which tends to slow the coarsening and the field gradient which tend to accelerate
the process.

\subsection{Goal and Results}

To summarize, the toy model (\ref{deftoy}) is thus interesting 
from several points of view. For the statics, it is directly related
to more complex problems ( directed polymer, burgers turbulence
and KPZ equation) in some particular limits, as described above; 
on the other hand, it represents in itself 
a `toy' disordered system that presents many
properties of more complex models, such as the presence of 
metastable states and the identities of the statistical tilt symmetry. 

Up to now, the landscape $U(x)$ of the toy model
has been characterized by two kinds of approach.
There has been on one side some studies on the absolute minimum
of the toy model, either in probability theory \cite{groeneboom}
 and in the context of the Burgers equation \cite{frachebourgmartin},
and on the other side a study \cite{cavagna} on the 
statistics of all extrema of the model (\ref{deftoy}), 
but that does not distinguish between the
important ones (the deep ones)
 and the large number of unimportant small scale extrema, nor
does it yield any information on joint distributions. 
None of these studies addresses the question of the joint distributions in energy and position
of the set of all deep extrema, neither the one relevant for the statics (several minima) 
and even less the one relevant for the dynamics (maxima, minima and barriers).

The goal of this paper is to show that the renormalization method can be
generalized to solve a broader class of one dimensional landscape than the
previously solved Brownian landscape \cite{dsfrg,us_long} and that it is
of deeper fundamental interest as an alternative to more conventional 
probability theory techniques. The class explicitly solved here includes the
toy model (\ref{deftoy}) landscape.
We formulate the RSRG for a general one-dimensional landscape as an exact method
to obtain the joint probability
distribution, in total number, position and value of energy
on the subset of the important deep extrema.
More precisely, the renormalized landscape at scale $\Gamma$ 
is defined by the subset of extrema of the initial landscape
which have survived after the erasing of all barriers
smaller than $\Gamma$. This is equivalent to require
that in a segment between a maximum $U(M)$ and the next minimum $U(m)$
of the renormalized landscape, any two points $(x,y)$
with $M<x<y<m$ must satisfy the constraint 
that the energy difference $U(y)-U(x)$ is smaller than $\Gamma$
 (and the symmetrical condition for a
 segment between a minimum and the next maximum). 
The joint probability distribution of this subset of important extrema
contains a tremendous amount of information, including
 in particular
the distribution of absolute extrema, of secondary extrema,
 of largest barriers etc...
Of course the general case is intractable analytically,
but our aim is to identify solvable models and universality classes,
 given by some continuum limits.

To be more specific, we will consider
the continuum models  
where the potential $U(x)$ is a stochastic process defined by the Langevin equation
\begin{eqnarray}
\frac{dU(x)}{dx} = F[U(x),x] + \eta(x)
\label{langevin} 
\end{eqnarray}
$\eta(x)$ being a delta correlated white noise.
The toy model (\ref{deftoy}) corresponds to the case where 
the force doesn't depend on $U(x)$ and is linear in $x$
\begin{eqnarray}
F[U(x),x] = F[x] = \mu x 
\label{forcetoy}
\end{eqnarray}
Another cases of interest are the stationary landscapes, 
where the force in the Langevin equation (\ref{langevin})
is independent of the space $x$ 
\begin{eqnarray}
F[U,x] = F[U] = - \frac{d W[U]}{d U}
\label{defstatio}
\end{eqnarray}
The case $F[U]=0$ is the pure Brownian landscape,
and the case $F[U]=F>0$ is the biased Brownian landscape.
These are the only two cases where the RSRG has been applied up to now
\cite{dsfrg,us_long}, but we will show here that it generalizes to
include any $F[U]$. Note that
solving the landscape problem here also amounts to solve the problem of
the joint distribution of extremal points in the trajectory of a particle 
in a one dimensional force field $F$ where $x$ is interpreted as "time" and
$U$ as the position of the particle.
For all cases (\ref{langevin}), the Markovian character 
of the Langevin equation
implies that the landscape measure 
can be written as product of blocks measures which satisfy 
closed real space renormalization equations.
For the specific cases of the toy model (\ref{forcetoy}) and of stationary
landscapes (\ref{defstatio}), we obtain here explicit expressions for the
blocks measures. This is achieved by two complementary methods
very different in spirit. On one hand, for the above continuum models
we are able to directly compute the blocks using some constrained path integrals.
On the other hand, we can also start from the RSRG equations for the blocks
and assume some Ansatz which solves these equations and obtain, 
interestingly via the Liouville equation, analytical forms for the blocks.
As is customary in searching for RG fixed points, 
this can be done without specifying the initial landscape,
and as a result the solutions look a priori more general. We show 
that these do contain the universality classes 
corresponding to (\ref{langevin}), with a result which
coincide exactly with one obtained by the direct path integral method. Our study thus
makes more precise the interesting relations between the powerful RSRG method and 
more conventional probabilistic approach using stochastic calculus. 

For the case of the toy model, the full solution for the joint 
probability measure of the renormalized landscape 
is obtained in this way explicitly in terms of Airy functions.
We then use this exact measure of the renormalized landscape
to study in details the equilibrium and the non-equilibrium
dynamics of the toy model (\ref{deftoy}) and give some
results for the RFIM in a field gradient.

For the equilibrium, we study the statistical properties of
the absolute minimum which dominates at zero temperature.
We obtain simple exact expressions for the moments of the
position of a single particle at equilibrium and 
and compare with the results of approximate methods, namely the
gaussian variational replica study 
\cite{mezardparisi92,engel} and of the vector breaking 
of replica symmetry \cite{dotsenko_mezard}.
We also compute the properties of
configurations with nearly degenerate minima which govern
the thermal fluctuations at very low-temperature, and we
prove that the rare configurations presenting two nearly degenerate minima
do not only give a finite contribution to (\ref{second}),
but actually are entirely responsible of the full result (\ref{second}).

For the dynamics of a single particle
in the toy energy landscape (e.g. a single RFIM domain wall),
we compute the distribution over samples of the equilibration time
$t_{eq}$ or equivalently the distribution of the 
largest barrier $\Gamma_{eq}=T \ln t_{eq}$ in the system. 
Our results are expressed in terms of the dimensionless scaling parameter  
\begin{eqnarray}
\gamma_{eq} = \left(\frac{\mu}{2}  \right)^{1/3} T \ln t_{eq}  
\end{eqnarray}
and we give in particular the two asymptotic behaviors
 of the distribution for $\gamma_{eq}$ small and large. 
We characterize the rare configurations presenting an
anomalously large equilibration time which govern the
long-time dynamics. We also compute local properties 
of the renormalized landscape, such as the probability
density of a minimum at scale $\Gamma=T \ln t$, and the 
the disorder averaged diffusion front for a single particle, which interpolates between
the Kesten distribution of the Sinai model at short rescaled time
and the reaching of equilibrium at long rescaled time. 

These quantities can be translated into results for
 observables of the RFIM in
a field gradient, assuming convergence to the full state where each maximum of the renormalized landscape contains an A domain and each minimum a B
domain. 
This convergence
was proved in \cite{us_long_rf} and also holds here in range of parameters of interest (small $\mu$, large time $t$) and with a random initial condition. As time $\Gamma = T \ln t$ increases, decimations occur
 corresponding to pairs of $A-B$ domains annihilating. Thus initially,
many domain coarsening take place. Depending
on the scaling parameter:
\begin{eqnarray}
\gamma_{eq0} = 2 J \left(\frac{\mu}{2}  \right)^{1/3} 
\end{eqnarray}
the equilibrium state, reached at time $T \ln t_{eq0}=2 J$ 
will contain either a single domain wall $B$ (for $\gamma_{eq0} \to + \infty$) or several additional
pairs $A-B$ (infinitely many as $\gamma_{eq0} \to 0$). The behaviors of the sample to sample
probabilities that a single domain wall $B$ remains is obtained. 
A formal solution for the
total magnetization is easily written, and an 
explicit calculation is performed for
large $\gamma_{eq0}$ where a single domain wall remains. Indeed in that limit 
the total magnetization is simply $M = 2 x_B$
where $x_B$ is the absolute minimum of the toy landscape.

The paper is organized as follows.
The Section \ref{generalrg} is devoted
 to the general definitions and properties of the renormalization 
procedure for an arbitrary one-dimensional landscape. In Section \ref{explicitsolu},
we compute the probability measure of the renormalized landscape for
a large class of models. We then study in details
 the toy model : we construct
the probability measure for the renormalized landscape 
in Section \ref{landscapetoy}, we study its equilibrium
properties in Section \ref{equilibriumtoy}, and 
the non-equilibrium dynamics in Section \ref{dynamicstoy}.
Some results for the RFIM in a field gradient are also discussed
in Section \ref{dynamicstoy}.
The Section \ref{statiolandscape} contains our results for the case of stationary landscapes. 
Finally, to facilitate reading of the paper, we give the more detailed definitions
and derivations in the several appendices.

\section{Renormalization procedure for an arbitrary 1D landscape}

\label{generalrg}

This section contains the precise definitions of the measure
for the initial landscape and for the renormalized landscape. 
The probability measure of the
renormalized landscape satisfies by construction a set of RG 
equations for the joint distribution of extrema. 

\subsection{Initial landscape}

\label{generalinitial}

The general model studied in this paper is the following. We consider one dimensional
random landscapes $U(x)$ which for definiteness in this section
are continuous in the energy $U$, discrete in space $x$ and defined
on a finite interval $x=1,2,...N$.
A realization of the random landscape is thus simply given
by a set of random variables $U_x$, $x=1,2,...N$ and the
original problem is defined by some probability measure noted

\begin{eqnarray}
{\cal P}[U] DU \equiv {\cal P}(U_1,...U_N) ~~ dU_1 ... dU_N
\label{initialmeasure}
\end{eqnarray}
The normalization with the fixed boundary conditions $U(x_1=x_L)=U_L$ and $U(x_N=x_R)=U_R$ reads
\begin{eqnarray}
Z(U_L,x_L;U_R,x_R) = \int {\cal P}(U_L,U_2,...U_{N-1},U_R) ~~ dU_2 ... dU_{N-1}
\label{normameasure}
\end{eqnarray}
We are interested in the statistics of the extrema of this landscape. 
For definiteness, we choose here the convention that
there are infinite potential outside the interval, i.e. $U(x=1^-)=U(x=N^+)=+ \infty$,
and that $N=2K-1$ is odd. We then define the probability
 $N_{\text{init}}^{(2n)} \left(U_L ; u_1, x_1 ; 
\ldots u_{2n-1}, x_{2n-1} ; x_R, U_R \right)$ that
the initial landscape has
$n$ local minima situated at positions $x_1,x_3..x_{2n-1}$ 
 with the potentials
$u_1,...u_{2n-1}$ and $(n-1)$ local maxima situated 
at positions $x_2,x_4..x_{2n-2}$ with potentials
$u_2,...u_{2n-2}$. They are of course alternated $x_1 < x_2 .. <x_{2n-1}$
and the values $x_1=x_L$ and $x_{2n-1}=x_R=2K-1$ are allowed.
An explicit and more precise definition of these objects 
is given in Appendix \ref{calculsn2}, with a derivation of their
normalization property 
\begin{eqnarray}
&& Z(U_L,x_L;U_R,x_R) =  \sum_{n=1}^{K}  \int du_1 du_2 ... du_{2n-1} \theta( U_L-u_1) \theta( U_R-u_{2n-1}) \prod_{i=1,n-1} \theta(u_{2i}-u_{2i-1} ) \theta(u_{2i}-u_{2i+1} )  \nonumber \\
&& \sum_{ x_L \leq x_1 < x_2 <x_3 <x_{2n-1} \leq x_R}  
 N_{\text{init}}^{(2n)} \left(U_L,x_L ; u_1, x_1 ; \ldots u_{2n-1}, x_{2n-1} ; U_R,x_R \right)
\label{normainitiale}
\end{eqnarray}
The objects $N_{\text{init}}^{(2n)}$ are usually extremely complicated and contain 
a lot of irrelevant information. To single out the important, deep extrema we now
define the renormalized landscape where all small barriers have been eliminated.

\subsection{Construction of the renormalized landscape}

\label{generalgamma}

The renormalized landscape at scale $\Gamma$ is defined 
by the list of positions and energies of
a {\it subset} $S_\Gamma$ of extrema of the initial landscape. 
For $\Gamma =0$, $S_0$ is the set of all extrema of the
initial landscape. For $\Gamma>0$ it is the subset of extrema
that survive when all barriers smaller than $\Gamma$ in the system
have been erased. This is shown in Fig 1. 
This subset $S_\Gamma$ is again made of alternating minima and maxima.
It is convenient to call descending bonds
the segments between a maximum and the next minimum both in $S_\Gamma$
 (and respectively ascending bonds the segments
between a minimum and the next maximum). 
The constraint obeyed by the subset is 
that any two points $(i,j)$ on a bond $[x , y]$ with $x<i< j < y$
 must satisfy 
that their energy difference $U_j - U_i$ is smaller 
than $\Gamma$ if the bond is descending,
and symmetrically, the energy difference $U_i - U_j$ must be smaller than $\Gamma$ if the bond is ascending.

\begin{figure}[thb]

\centerline{\fig{12cm}{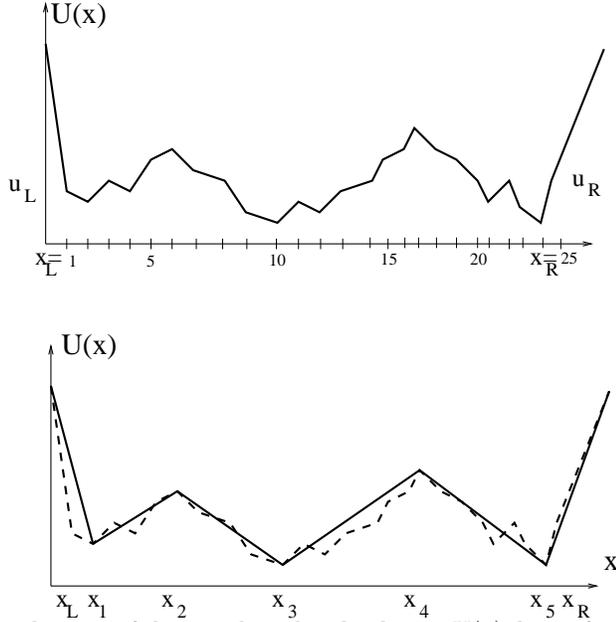}} 
\caption{ {\it top} One realization of the initial random landscape $U(x)$
drawn from the probability distribution ${\cal P}[U]$ with $N=25$.
It contributes to $N_{\text{init}}^{(2n)} \left(x_L,u_L ; u_1, x_1 ; 
\ldots u_{2n-1}, x_{2n-1} ; x_R, u_R \right)$ with $n=6$ and 
$(x_1,..x_{11})=(2,3,4,6,10,11,12,17,21,22,24)$ since only $2n-1=11$
points of the initial landscape are actually extrema. {\it bottom}
The corresponding realization of the renormalized landscape where
all barriers smaller than some given $\Gamma$ have been removed. It
contributes to  $N_{\Gamma}^{(2n')}
\left(x_L,u_L ; u_1, x_1 ; \ldots u_{2n'-1}, x_{2n'-1} ; x_R, u_R \right)$
with $n'=3$. As in the top figure, the straight lines 
are drawn only for convenience, the landscape being
defined solely by the set of $(x_i,u_i)$, $i=1,..,2n'-1=5$ of
remaining extrema, 
together with the right and left boundary points. \label{fig1} } 

\end{figure}

The main object of interest is the probability measure of the renormalized landscape. We consider the probability
 $N_{\Gamma} \left(x_L,u_L ; u_1, x_1 ;
 \ldots u_{2n-1}, x_{2n-1} ; x_R, u_R \right)$
 that the renormalized landscape has
$n$ local minima situated at positions $x_1,x_3..x_{2n-1}$ 
 with the potentials
$u_1,...u_{2n-1}$ and $(n-1)$ local maxima situated 
at positions $x_2,x_4..x_{2n-2}$ with potentials
$u_2,...u_{2n-2}$. Again, more precise and explicit definitions are given in Appendix 
\ref{calculsn2}, with a derivation of the normalization 
\begin{eqnarray}
&& Z(U_L,x_L;U_R,x_R) =  \sum_{n=1}^{K} \int du_1 ... du_{2n-1} \theta( U_L-u_1) \theta( U_R-u_{2n-1})\prod_{i=1,n-1} \theta(u_{2i}-u_{2i-1}-\Gamma ) \theta(u_{2i}-u_{2i+1}-\Gamma ) \nonumber \\
&& \sum_{ x_L \leq x_1 < x_2 <x_3 <x_{2n-1} \leq x_R}  
 N_{\Gamma}^{(2n)} \left(U_L,x_L ; u_1, x_1 ; \ldots u_{2n-1}, x_{2n-1} ; U_R,x_R \right)
\label{normagamma}
\end{eqnarray}
 The normalization of the measure
for the renormalized landscape
thus coincides
with the normalization of the measure for the initial landscape
(\ref{normameasure},\ref{normainitiale}).

\begin{figure}[thb]

\centerline{\fig{6cm}{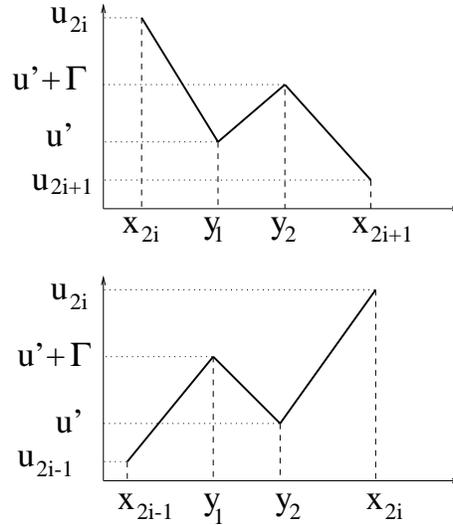}} 
\caption{ Terms in the RSRG equation (\ref{rsrg}) corresponding to the elimination
of two local extrema separated by a potential difference $\Gamma$, for 
a descending bond (top) and ascending (bottom) respectively \label{fig2} } 

\end{figure}

The $N_{\Gamma}^{(2n)}$ satisfy the RG equation (see Appendix \ref{calculsn2})
\begin{eqnarray}
&& \partial_{\Gamma} N_{\Gamma}^{(2n)}
\left(x_L,u_L ; u_1, x_1 ; \ldots u_{2n-1}, x_{2n-1} ; x_R, u_R \right) 
\nonumber  \\
= && \int_{x_L}^{x_1} dy_1 \int_{y_1}^{x_1} dy_2 \int_{u_1}^{+\infty} du'
N_{\Gamma}^{(2n+2)}
\left(x_L,u_L ; u' , y_1 ; u'+\Gamma , y_2 ; u_1, x_1 ; \ldots u_{2n-1}, x_{2n-1} ; x_R, u_R \right)  \nonumber \\
&& +\sum_{i=1}^{n-1} 
\int_{x_{2i}}^{x_{2i+1}} dy_1 \int_{y_1}^{x_{2i+1}} dy_2 
\int_{u_{2i+1}}^{u_{2i}-\Gamma} du'
N_{\Gamma}^{(2n+2)}
\left(x_L,u_L ;  \ldots ; u_{2i} , x_{2i}  ; u' , y_1 ; u'+\Gamma , y_2 ;
u_{2i+1} , x_{2i+1} ;\ldots ; x_R, u_R \right)  \nonumber \\
&& +\sum_{i=1}^{n-1} 
\int_{x_{2i-1}}^{x_{2i}} dy_1 \int_{y_1}^{x_{2i}} dy_2 
\int_{u_{2i-1}}^{u_{2i}-\Gamma} du'
N_{\Gamma}^{(2n+2)}
\left(x_L,u_L ;  \ldots ; u_{2i-1} , x_{2i-1}  ; u'+\Gamma , y_1 ; u' , y_2 ;
u_{2i} , x_{2i} ;\ldots ; x_R, u_R \right)  \nonumber \\
&& +\int_{x_{2n-1}}^{x_R} dy_1 \int_{y_1}^{x_R} dy_2 \int_{u_{2n-1}}^{+\infty} du'
N_{\Gamma}^{(2n+2)}
\left(x_L,u_L ; u_1, x_1 ; \ldots u_{2n-1}, x_{2n-1} ;
 u'+\Gamma , y_1 ; u', y_2 ; x_R, u_R \right) 
\label{rsrg}
\end{eqnarray}
where, for convenience and future applications to
continuum limits, we have used the integral symbol instead of the sum,
i.e. $\int_{y}^{x} dz$ instead of $\sum_{z=y+1}^{x-1}$.
These RSRG equation describe the decimation
upon increasing $\Gamma$ and are thus a generalization to an arbitrary landscape
of the standard RSRG equations considered in the case of a Brownian landscape
(see the formulae (197) and the figure 8 in Ref \cite{us_long}).
It simply expresses that as $\Gamma$ is increased,
pairs of extrema separated by an energy barrier $\Gamma$ are eliminated. 
Each term in (\ref{rsrg}) corresponds to possible positions of this pair, as represented in Fig
\ref{fig2}, the first and last one corresponding to edge bonds.

The aim of this paper is to find solvable cases where analytical solutions of
(\ref{rsrg}) are possible. The general case with arbitrary correlations in
the original landscape can only be studied by numerical 
implementations of the decimation procedure. This has been done, for instance, in
the case of a logarithmically correlated landscape in \cite{castillo}.
However, for the initial landscapes that are defined by an initial local measure, 
we expect that the measure for the renormalized landscape can be decomposed 
as a product of blocks associated to bonds as we now describe. 

\subsection{Markovian landscapes and Block-product measures}

From now on, we will consider ``Markovian landscapes",
corresponding to the cases where the probability measure of the
initial landscape (\ref{initialmeasure})
is a product of block measures of the form:
\begin{eqnarray}
{\cal P}[U] = \Phi_0(U_1)
\Phi_N(U_N) \prod_{i=1}^{N-1} \exp( -  S_i[U_{i+1}, U_{i}] ) 
\end{eqnarray}
It follows immediately that the probabilities of the renormalized landscape
can be written as a product of block probabilities which depend only on the endpoints of the bonds
block, all the information inside the bonds have been in a sense integrated out (\ref{explidecoupled}) :

\begin{eqnarray}
&& N_{\Gamma}^{(2n)}
\left(x_L,u_L ; u_1, x_1 ; \ldots u_{2n-1}, x_{2n-1} ; x_R, u_R \right)   
 =
E^+_{\Gamma} ( x_L,u_L ; u_1, x_1 )
B^-_{\Gamma} ( u_1, x_1 ;  u_2, x_2 )  \nonumber \\
&& B^+_{\Gamma} ( u_2, x_2 ;  u_3, x_3 )  \ldots 
 B^+_{\Gamma} ( u_{2n-2}, x_{2n-2} ;  u_{2n-1}, x_{2n-1} )
E^-_{\Gamma} (  u_{2n-1}, x_{2n-1} ; x_R, u_R  ) 
\label{decoupled}
\end{eqnarray}
 Note that 
the blocks $B^\pm_{\Gamma} ( u_2, x_2 ;  u_3, x_3 )$ themselves do not in general 
satisfy any normalization. The only constraint is the global normalization:
\begin{eqnarray}
Z(u_L,x_L;u_R,x_R) = \sum_{n=1}^K Tr(E^+ (B^- B^+)^{n-1} E^-)
\end{eqnarray}
where we have used an obvious matrix notation with respect to $(x,u)$. Each
term in this sum, $N^{(2n)} = Tr(E^+ (B^- B^+)^{n-1} E^-)$, represents the
probability that the renormalized landscape has exactly $(2 n -1)$ extrema.

The insertion of the decoupled form (\ref{decoupled})
into the RSRG equation (\ref{rsrg}) 
yields the following RG equations for the 
block measures 
\begin{eqnarray}
&& \partial_{\Gamma} E^{+}_{\Gamma} ( x_L,u_L ;  u, x ) \nonumber  \\
= && \int_{x_L}^{x} dy_1 \int_{y_1}^{x} dy_2 \int_{u}^{+\infty} du'
E^{+}_{\Gamma} ( x_L,u_L ;  u', y_1 )
B^{-}_{\Gamma} (  u', y_1 ; u'+\Gamma,y_2 )
B^{+}_{\Gamma} ( u'+\Gamma,y_2 ; u,x ) \\
&& \partial_{\Gamma} B^{-}_{\Gamma} ( u,x ;  u'', x'') \nonumber  \\
= &&
\int_{x}^{x''} dy_1 \int_{y_1}^{x''} dy_2 
\int_{u}^{u''-\Gamma} du'
B^{-}_{\Gamma} ( u,x ;  u' +\Gamma, y_1 )
B^{+}_{\Gamma} (  u'+\Gamma, y_1 ; u',y_2 )
B^{-}_{\Gamma} ( u',y_2 ; u'',x'' ) \\
&& \partial_{\Gamma} B^{+}_{\Gamma} ( u,x ;  u'', x'') \nonumber  \\
= &&
\int_{x}^{x''} dy_1 \int_{y_1}^{x''} dy_2 
\int_{u''}^{u-\Gamma} du'
B^{+}_{\Gamma} ( u,x ;  u', y_1 )
B^{-}_{\Gamma} (  u', y_1 ; u'+\Gamma,y_2 )
B^{+}_{\Gamma} ( u'+\Gamma,y_2 ; u'',x'' ) \\
&& \partial_{\Gamma} E^{-}_{\Gamma} (   u, x ; x_R, u_R ) \nonumber  \\
= && \int_{x_L}^{x} dy_1 \int_{y_1}^{x} dy_2 \int_{u}^{+\infty} du'
B^{-}_{\Gamma} (  u, x ; u'+\Gamma,y_1 )
B^{+}_{\Gamma} ( u'+\Gamma,y_1 ; u',y_2 ) 
E^{-}_{\Gamma} ( u',y_2 ; x_R, u_R )
\label{eqrgb}
\end{eqnarray}
Again note that the factorized form and these RG equations hold either for discrete space
(with the abovementioned correspondence between the integral symbol and the sum) 
as well as in the continuum limit where the space cutoff goes to zero.

\section{Explicit Solutions for a class of models}

\label{explicitsolu}

\subsection{Soluble models}

\label{solvable}

To be more specific we now consider
the continuum models of Markovian landscapes 
where the potential $U(x)$ is a stochastic process defined by the Langevin equation
(\ref{langevin}).
This process corresponds to functional integration with the Fokker-Planck measure
\begin{eqnarray}
{\cal P}[U] \sim  \exp( - \int dx ( \frac{1}{4} (\frac{dU(x)}{dx} - F[U(x),x])^2 +
\frac{1}{2} \partial_U F[U(x),x] )
\label{fokker-planck}
\end{eqnarray}

In the following, we will give an explicit solution
for the landscape renormalization for all the cases
where {\it the Fokker-Planck measure
can be rewritten as a Schrodinger measure }
\begin{eqnarray}
{\cal P}[U] \sim  exp( - \int dx ( \frac{1}{4} (\frac{dU(x)}{dx})^2 +
V[U(x)] )
\label{schrodinger}
\end{eqnarray}
{\it up to boundaries terms}. The two important cases are 
the toy model defined in (\ref{deftoy})
and the stationary landscapes (\ref{defstatio}).

\subsubsection{Toy model}

The toy model (\ref{deftoy}) 
corresponds to the case where the force in the Langevin
equation (\ref{langevin}) is given by (\ref{forcetoy}).
 The Fokker-Planck measure for the toy model can be be rewritten
as a Schrodinger measure
up to boundary terms as follows
\begin{eqnarray}
\exp(- \int_{x_1}^{x_2} dx ( \frac{1}{4  } (\frac{dU}{dx} - \mu x)^2 ) = 
e^{ h(u_1,x_1)-h(u_2,x_2) } 
\exp( - \int_{x_1}^{x_2} dx 
( \frac{1}{4 } (\frac{dU}{dx})^2 + \frac{1}{2} \mu U ))
\label{transfotoy}
\end{eqnarray}
where the boundary terms reads
\begin{eqnarray}
h(u,x) = -\frac{\mu}{2  } x u  + \frac{\mu^2}{12 } x^3 
\label{defhtoy}
\end{eqnarray}
The associated Schrodinger potential is linear in $U$
\begin{eqnarray}
V[U]= \frac{1}{2 } \mu U
\label{potlinear}
\end{eqnarray}

\subsubsection{Stationary landscapes}

The stationary landscapes 
 correspond to the cases where the force in the Langevin equation (\ref{langevin})
is independent of space (\ref{defstatio}).
The transformation from Fokker-Planck measure to Schrodinger measure
up to boundary terms is as follows:
\begin{eqnarray}
&& \exp(- \int_{x_1}^{x_2} dx ( \frac{1}{4} (\frac{dU}{dx} - F[U])^2 +
\frac{1}{2} F'[U] ))  \nonumber \\
&& = 
e^{  \frac{1}{2} (W[U(x_1)] -  W[U(x_2)]) } 
\exp( - \int_{x_1}^{x_2} dx ( \frac{1}{4} (\frac{dU}{dx})^2 + \frac{1}{4} F[U]^2 + 
\frac{1}{2} F'[U] ))
\label{transfostatio}
\end{eqnarray}
The associated Schrodinger potential reads
\begin{eqnarray}
V[U]=\frac{1}{4} F[U]^2 + \frac{1}{2} F'[U]=\frac{1}{4} W'[U]^2 - \frac{1}{2} W''[U] 
\label{potstatio}
\end{eqnarray}

\subsection{Properties of the blocks for a Schrodinger measure}

For a Schrodinger measure (\ref{schrodinger}), the important simplification
comes from the translation invariance of the blocks 
\begin{eqnarray}
&&  B^{\pm}_{\Gamma} ( u, x ;  u'', x'')
= B^{\pm}_{\Gamma} ( u ;  u'' ; x''-x )  \\
&& E^{+}_{\Gamma} ( u_L,x_L ;  u, x ) = E^{+}_{\Gamma}(u_L,u, x-x_L)  \\
&& E^{-}_{\Gamma} (    u, x ; u_R,x_R) = E^{+}_{\Gamma}(u,u_R, x_R-x)
\end{eqnarray}
which will allow us to obtain explicit expressions for the Laplace transforms
\begin{eqnarray}
\label{defblaplace}
B^{\pm}_{\Gamma} (u ; u''; p) 
&& = \int_0^{+\infty} dx B^{\pm}_{\Gamma} (u ;  u'' ; x) \\
E^{\pm}_{\Gamma} (u ; u''; p) 
&& = \int_0^{+\infty} dx E^{\pm}_{\Gamma} (u ;  u'' ; x)
\nonumber 
\end{eqnarray}
The symmetry $x \to -x$ moreover implies:
\begin{eqnarray}
 B^{+}_{\Gamma} ( u , x ;  u'', x'') 
= B^{-}_{\Gamma} ( u , -x ; u'' , -x'')
\end{eqnarray}
so that in Laplace we have the simple relations
\begin{eqnarray}
 B^{+}_{\Gamma} ( u ;  u'' ; p) 
= B^{-}_{\Gamma} ( u'' ;  u ; p) \\
 E^{+}_{\Gamma} ( u ;  u'' ; p) 
= E^{-}_{\Gamma} ( u'' ;  u ; p)
\label{symmbpm}
\end{eqnarray}
To determine these functions, we now present two methods
which are rather different in spirit and complementary.

\subsection{Explicit solution via a path-integral approach}

For the Schrodinger measure, the block $B^-_{\Gamma} $
can be represented by a path-integral 
\begin{eqnarray}
B^-_{\Gamma} ( u_0, x_0 ;  u, x ) = 
\int_{U_{x_0}=u_0}^{U_{x}=u} DU 
exp( - \int_{x_0}^x dy ( \frac{1}{4} (\frac{dU}{dy})^2 + V[U] )
\Theta_\Gamma^-(x_0,x;U)
\label{pathbm}
\end{eqnarray}
which is defined for $u \ge u_0 + \Gamma$ and where the symbol
$\Theta_\Gamma^-(x_0,x;U)$ constraints the paths to
remain in the interval $]u_0,u[$ for $x_0<y<x$ and
to not perform any returns of more than $\Gamma$. 
As shown in Appendix (\ref{pathcomputation}),
this constrained path-integral can be expressed in
Laplace transform with respect to $(x-x_0)$ as
 \begin{eqnarray}
 B^-_{\Gamma} ( u_0, u,p) \equiv \int_{x_0}^{+\infty} dx
e^{-px} B^-_{\Gamma} ( u_0, x_0 ;  u, x ) && = \frac{1}{K(u_0,u_0+\Gamma,p)} 
\exp( - \int_{u_0 + \Gamma}^{u} du' \partial_2 \ln K(u' - \Gamma,u' ,p) ) \\
&& = \frac{1}{K(u- \Gamma,u,p)} 
\exp( \int_{u_0 }^{u- \Gamma} dv \partial_1 \ln K(v,v+ \Gamma ,p) ) 
\label{resbmpath}
\end{eqnarray}
(the two expressions are equivalent via an integration by part)
in terms of the function
\begin{eqnarray}
K(u,v,p) = \frac{1}{w(p)} (\phi_1(u,p) \phi_2(v,p) 
- \phi_2(u,p) \phi_1(v,p))
\label{defK}
\end{eqnarray}
where $\phi_1(u,p)$
and $\phi_2(u,p)$ are
two linearly independent solutions  of the equation
\begin{eqnarray}
\partial^2_u F - V(u) F = p F
\label{eqF}
\end{eqnarray}
of wronskian ( independent of $u$ )
\begin{eqnarray}
w(p)=\phi_1(u,p) \phi_2'(u,p) - \phi_2(u,p) \phi_1'(u,p)
\end{eqnarray}
Similarly for the edges, the final result reads (
see Appendix \ref{pathcomputation}) 
\begin{eqnarray}
 E^-_{\Gamma} (u, u_R,p) =
\exp(  \int_{u}^{u_R} \partial_1 \ln K(u' , u' + \Gamma,p) du')
\label{resempath}
\end{eqnarray}

\subsection{Explicit solution from the RSRG equations}

The RG equations for the blocks $B$ (\ref{eqrgb}) are simpler in Laplace
variables :
\begin{eqnarray}
&& \partial_{\Gamma} B^{-}_{\Gamma} ( u ;  u'' ; p) 
= \int_{u}^{u''-\Gamma} du'
B^{-}_{\Gamma} ( u ;  u' +\Gamma ; p )
B^{+}_{\Gamma} ( u'+\Gamma ; u' ; p )
B^{-}_{\Gamma} ( u' ; u'' ; p ) \\
&& \partial_{\Gamma} B^{+}_{\Gamma} ( u ;  u'' ; p) =
\int_{u''}^{u-\Gamma} du'
B^{+}_{\Gamma} ( u ;  u' ; p)
B^{-}_{\Gamma} ( u' ; u'+\Gamma ; p)
B^{+}_{\Gamma} ( u'+\Gamma ; u'' ; p)
\end{eqnarray}
Using the symmetry (\ref{symmbpm}), we obtain that
 $B^{-}_{\Gamma} ( u'' ;  u ; p)$ satisfies the closed RG equation
\begin{eqnarray}
\partial_{\Gamma} B^{-}_{\Gamma} ( u ;  u'' ; p) 
= \int_{u}^{u''-\Gamma} du'
B^{-}_{\Gamma} ( u ;  u' + \Gamma ; p)
B^{-}_{\Gamma} ( u' ; u'+ \Gamma ; p)
B^{-}_{\Gamma} ( u' ; u'' ; p)
\label{closed}  
\end{eqnarray}
and the edges satisfy
\begin{eqnarray}
 \partial_{\Gamma} E^{-}_{\Gamma} (u,u_R, p) = \int_{u}^{+\infty} du'
B^{-}_{\Gamma} ( u ; u' + \Gamma ; p )
B^{-}_{\Gamma} ( u' ; u' + \Gamma ; p )
E^{-}_{\Gamma} (u',u_R,p)
\end{eqnarray}

As shown in Appendix (\ref{appliouville}), 
a class of solutions of the RSRG equation (\ref{closed})
can be found with the factorized Ansatz
\begin{eqnarray}
B^{-}_{\Gamma} ( u ; u'' ; p) = A^L_\Gamma(u,p)  A^R_\Gamma(u'',p)
\label{formdec}
\end{eqnarray}
provided that the function $\Phi$ defined by
\begin{eqnarray}
\partial_1 \partial_2 \Phi(u , u + \Gamma) 
= B^{-}_{\Gamma} ( u ; u + \Gamma ; p )^2
\label{carre}
\end{eqnarray}
satisfies the following Liouville equation
\begin{eqnarray}
\partial_1 \partial_2 \Phi(u_1, u_2) = \exp(- 2 \Phi(u_1 , u_2))
\label{liouville}
\end{eqnarray}
The general solution, discussed in more details in Appendix (\ref{appliouville}),
can be written
\begin{eqnarray}
&& \Phi(u_1, u_2) = \ln 
\left[  \frac{1}{\sqrt{w_L w_R}} (\psi_L^{(1)}(u_1) \psi_R^{(2)}(u_2) - 
\psi_L^{(2)}(u_1) \psi_R^{(1)}(u_2) )   \right]
\label{parawronsk}
\end{eqnarray}
where $\psi_L^{(1,2)}(u)$ and $\psi_R^{(1,2)}(u)$ are respectively
two pairs of two linearly independent solutions of the two Schrodinger
equations:
\begin{eqnarray}
&& \psi_L''(u) = V_L(u,p) \psi_L(u) \\
&& \psi_R''(u) = V_R(u,p) \psi_R(u)
\end{eqnarray}
where $V_L(u,p)$ and $V_R(u,p)$ are 
arbitrary potentials, and where the constants $
w_{L,R}$ are the two wronskian, i.e
$w_{L,R} = \psi_{L,R}^{(1)}(u) {\psi_{L,R}^{(2)}}'(u) -
\psi_{L,R}^{(2)}(u) {\psi_{L,R}^{(1)}}'(u) $.

The results of the path-integral approach (\ref{pathbm})
are recovered with the following choice:
\begin{eqnarray}
V_L(u,p)= V_R(u,p) = V(u) + p 
 \label{choice}
\end{eqnarray}
with the identifications $\psi_L^{(1)}=\psi_R^{(1)}$,  $\psi_L^{(2)}=\psi_R^{(2)}$ and
\begin{eqnarray}
&& \Phi(u_1,u_2) = \ln K(u_1,u_2)
\end{eqnarray}

Thus by solving the RSRG equations with the following properties: 
(i) factorization in blocks (ii)
statistical translational invariance and parity of the blocks
 (iii) an additional
factorization property in Laplace variable (\ref{formdec}),
we have found general solutions involving two
arbitrary potentials $V_L(u,p)$ and $V_R(u,p)$.
This method does not use the knowledge of the initial landscape 
they correspond to. This is
a familiar feature of RG equations, fixed points can be found under symmetry
assumptions but finding the basin of attraction is harder.
We have showed that the special choice (\ref{choice})
corresponds to the case of initial Schrodinger measure (\ref{pathbm})
that we have independently solved via the path-integral method.
It would be of course of great interest to understand
whether other choices for the potentials
$V_L(u,p)$ and $V_R(u,p)$ yields sensible solutions
 for the landscape problem. However this analysis goes beyond
the scope of this paper and it is 
left as an open question for future investigation. Note also that
all formulae here and in Appendix (\ref{appliouville}) hold for
discrete space, under the same assumptions (defining Laplace transforms 
as $B^{\pm}(u,u'',p)=\sum_{x=1}^{+\infty} B^{\pm}(u,u'',x)$), and
one could also check whether discrete landscape could also be solved
by these methods.

Note finally that although fixed point (i.e. continuum) models can
alternatively be solved by the constrained path integral method, the RSRG is in
principle more powerful. Here for instance we only need assumptions (i) to (iii)
to hold only asymptotically at large $\Gamma$. In principle the full convergence to the fixed point
could be also studied, as in any RG method, and universal results established for
a broader set of discrete models within the basin of attraction of the
fixed points found here. 

In the following we restrict to the case of Schrodinger measures
corresponding to the choice (\ref{choice}) and use the explicit solutions
derived above to study the case of the toy model (\ref{deftoy})
in details and the case of stationary landscapes (\ref{defstatio}) more briefly.

\section{Statistics of extrema in the toy model}

\label{landscapetoy}

In this section, we apply the general construction of the probability
measure of renormalized landscape to the particular case 
of the toy model (\ref{deftoy}). We start by 
giving the expressions for the blocks before constructing the full measure. 

\subsection{Blocks of the toy model in terms of Airy functions}

We have seen in (\ref{transfotoy}) how the functional integrals for the process
$dU/dx =  \mu x  + \eta(x)$ could be recast 
{\it up to boundary terms} into functional
integrals of the Schrodinger-type.
The Fokker-Planck blocks thus read  (\ref{transfotoy})
\begin{eqnarray}
&& B^\pm_{\Gamma}( u_0, x_0 , u_1, x_1) =
e^{ h(u_0,x_0)- h(u_1,x_1)} 
\tilde{B}^\pm_{\Gamma}( u_0,  u_1, x_1-x_0)  \\
&& E^\pm_{\Gamma}( u_0,  u_1, x_1 -x_0 ) =
e^{ h(u_0,x_0)- h(u_1,x_1)} 
\tilde{E}^\pm_{\Gamma}( u_0, x_0 , u_1, x_1)  
\label{changetotilde}
\end{eqnarray}
in terms of the Schrodinger blocks $\tilde{B}$ $\tilde{E}$ given in Laplace
(\ref{resbmpath},\ref{resempath}) by
\begin{eqnarray}
&& \tilde{B}^-_{\Gamma} ( u_0, u,p) = \tilde{B}^+_{\Gamma} ( u, u_0,p)
=  \frac{1}{K(u- \Gamma,u,p)} 
\exp( \int_{u_0 }^{u- \Gamma} dv \partial_1 \ln K(v,v+ \Gamma ,p) ) \\
&& \tilde{E}^+_{\Gamma} (u_b,u,p) = \tilde{E}^-_{\Gamma} (u, u_b,p)
= \exp(  \int_{u}^{u_b} \partial_1 \ln K(u', u' + \Gamma,p) du')
\label{resblocktoy}
\end{eqnarray}
The function $K$ defined in (\ref{defK}) 
 now corresponds to the problem 
with the Schrodinger potential 
$V[U] = \frac{1}{2} \mu U$ (\ref{potlinear}).
Two independent solutions of the Schrodinger equation are the Airy functions
(\ref{eqAi}) 
\begin{eqnarray} 
&& \phi_1(u,p) = Ai(a u + b p)  \\
&& \phi_2(u,p ) =  Bi(a u + b p) 
\label{elemtoy}
\end{eqnarray}
where
\begin{eqnarray}
 a &&  = \left(\frac{\mu}{2 }\right)^{1/3}  \\
 b && = \frac{1}{ a^2}
\label{toyscale}
\end{eqnarray}
$a$ is the characteristic inverse energy, whereas $b$
represents the characteristic length scale of the toy model.
The wronskian of the two solutions (\ref{elemtoy}) reads
\begin{eqnarray}
w(p)=\frac{a}{\pi}
\end{eqnarray}
and thus the function $K$ reads(\ref{defK})
\begin{eqnarray}
K(u,v,p)=\frac{\pi}{a}
\left( Ai(a u + b p) Bi(a v + b p)
- Bi(a u + b p) Ai(a v + b p) \right)
\label{ktoy}
\end{eqnarray}
Although there are no obvious simple expressions for the integrals
involved in the blocks (\ref{resblocktoy}), in calculations 
it will be often useful to 
extract the leading behavior for $u<v$ by rewriting
\begin{eqnarray}
K(u,v,p)=\frac{\pi}{a}
 Ai(a u + b p) Bi(a v + b p) \left( 1
- \frac{Bi(a u + b p)}{Ai(a u + b p)} 
 \frac{ Ai(a v + b p) }{Bi(a v + b p)} \right)
\label{kfactor}
\end{eqnarray}
and
\begin{eqnarray}
\partial_1 \ln K(u,v,p)=
&& \partial_u \ln Ai(a u + b p) 
+ \partial_u \ln \left( 1
- \frac{Bi(a u + b p)}{Ai(a u + b p)} 
 \frac{ Ai(a v + b p) }{Bi(a v + b p)} \right) 
\label{separation}
\end{eqnarray}

\subsection{Normalization of the full Measure for the renormalized landscape }

As explained in Section (\ref{generalrg}), we
 consider left and right reflecting boundaries,
i.e. we impose $u(x_L)=u_L$ and $u(x_R)=u_R$ with $u(x_L^-)=+\infty$ and $u(x_R^+)=+\infty$. 
Then the normalization of the measure is (\ref{fokker-planck}) :

\begin{eqnarray}
Z(u_L, x_L;u_R, x_R) =
\int_{u(x_L)=u_L}^{u(x_R)=u_R} DU
e^{- \frac{1}{4} \int_{x_L}^{x_R} dx   \left(\frac{dU}{dx} - \mu x \right)^2 }
= \frac{1}{\sqrt{4 \pi (x_R-x_L)}}
e^{- \frac{(u_R - u_L - \frac{\mu}{2} x_R^2 + \frac{\mu}{2} x_L^2 )^2}{4 (x_R-x_L)}}
\label{normaztoy}
\end{eqnarray}

The decomposition into blocks at scale $\Gamma$ reads (\ref{decoupled})

\begin{eqnarray}
  Z(u_L,x_L,u_R,x_R)  = && \sum_{n=1}^{+\infty} 
\int \prod_{i=1}^{2n-1} du_i dx_i 
E^+_{\Gamma} ( u_L, x_L ; u_1, x_1 )
B^-_{\Gamma} ( u_1, x_1 ;  u_2, x_2 )  \ldots   \nonumber \\
&& B^+_{\Gamma} ( u_{2n-2}, x_{2n-2} ;  u_{2n-1}, x_{2n-1} )
E^-_{\Gamma} (  u_{2n-1}, x_{2n-1} ; u_R ,x_R ) \\
&& = e^{ h(u_L,x_L)- h(u_R,x_R)}
\sum_{n=1}^{+\infty} 
\int \prod_{i=1}^{2n-1} du_i dx_i
\tilde{E}^+_{\Gamma} ( u_L, u_1, x_1 -x_L )
\tilde{B}^-_{\Gamma} ( u_1,  u_2, x_2-x_1 )  \ldots \nonumber \\
&& \tilde{B}^+_{\Gamma} ( u_{2n-2},    u_{2n-1}, x_{2n-1}-x_{2n-2} )
\tilde{E}^-_{\Gamma} (  u_{2n-1},  u_R ,x_R-x_{2n-1} ) 
\label{normaz2}
\end{eqnarray}
i.e. the boundary terms involving the function $h$ 
in the transformation from Fokker-Planck to Schrodinger (\ref{transfotoy})
cancel between two successive blocks, and thus only survive
at the boundaries of the full system $(x_L,x_R)$.

\subsection{Alternative expression of the full measure using barriers}

The invariance by a global shift in the energy $u$ of the Fokker-Planck blocks 
\begin{eqnarray} 
&& B^-_{\Gamma} ( u_0, x_0 ;  u_1, x_1 ) 
= B^-_{\Gamma} (0 , x_0 ;  u_1 - u_0 , x_1 ) \\
&& B^+_{\Gamma} ( u_0, x_0 ;  u_1, x_1 ) 
= B^+_{\Gamma} (u_0 - u_1 , x_0 ;  0 , x_1 ) 
\label{globalshift}
\end{eqnarray}
translates into the following properties
for the Schrodinger blocks (\ref{changetotilde}) 
using the expression of the function h (\ref{defhtoy})
\begin{eqnarray}
 \tilde{B}^-_{\Gamma}( u_0,  u_1, x_1-x_0)
&&  =
e^{ h(0,x_0) - h(u_0,x_0)+ h(u_1,x_1) - h(u_1-u_0,x_1)} 
\tilde{B}^-_{\Gamma}( 0,  u_1-u_0, x_1-x_0) \nonumber  \\
&&  =
e^{ \frac{\mu}{2} x_0 u_0 - \frac{\mu}{2} x_1 u_0} 
\tilde{B}^-_{\Gamma}( 0,  u_1-u_0, x_1-x_0)  \\
 \tilde{B}^+_{\Gamma}( u_0,  u_1, x_1-x_0)
&&  =
e^{ h(u_0-u_1,x_0) - h(u_0,x_0)+ h(u_1,x_1)  - h(0,x_1)} 
\tilde{B}^+_{\Gamma}( u_0-u_1, 0, x_1-x_0)  \nonumber \\
&&  =
e^{ \frac{\mu}{2} x_0 u_1 - \frac{\mu}{2} x_1 u_1} 
\tilde{B}^+_{\Gamma}( u_0-u_1, 0, x_1-x_0)
 \end{eqnarray}
and similarly for the edge blocks $E$.
An alternative expression of (\ref{normaz2}) containing only
differences of energies is thus given by

\begin{eqnarray}
&&  Z(u_L,x_L,u_R,x_R)  =  
\sum_{n=1}^{+\infty} 
\int \prod_{i=1}^{2n-1} du_i dx_i
[ e^{ \frac{\mu^2}{12} x_L^3 -\frac{\mu}{2} x_L (u_L-u_1) } 
\tilde{E}^+_{\Gamma}( u_L-u_1, 0, x_1-x_L) ]  \nonumber \\
&& 
 \tilde{B}^-_{\Gamma}( 0,  u_2-u_1, x_2-x_1)
e^{ \frac{\mu}{2} x_{2} (u_3-u_1) } 
\tilde{B}^+_{\Gamma}( u_{2}-u_3, 0, x_3-x_{2})  \nonumber \\
&& 
 \tilde{B}^-_{\Gamma}( 0,  u_4-u_3, x_4-x_3)
e^{ \frac{\mu}{2} x_{4} (u_5-u_3) } 
\tilde{B}^+_{\Gamma}( u_{4}-u_5, 0, x_5-x_{4})
 \ldots     \nonumber  \\
&& \tilde{B}^-_{\Gamma}( 0,  u_{2n-2}-u_{2n-3}, x_{2n-2}-x_{2n-3})
e^{ \frac{\mu}{2} x_{2n-2} (u_{2n-1}-u_{2n-3}) } 
\tilde{B}^+_{\Gamma}( u_{2n-2}-u_{2n-1}, 0, x_{2n-1}-x_{2n-2}) 
\nonumber \\
&& [ e^{ + \frac{\mu}{2} x_R (u_R-u_{2n-1}) - \frac{\mu^2}{12} x_R^3 }
\tilde{E}^-_{\Gamma}( 0,  u_R-u_{2n-1}, x_R-x_{2n-1}) ]
\label{decompofixed}
\end{eqnarray}
where 
\begin{eqnarray}
&& \tilde{B}^-_{\Gamma} (0,F,p) = \tilde{B}^+_{\Gamma} (F,0,p) =
 \frac{1}{K(F- \Gamma,F,p)} 
\exp( \int_{0 }^{F - \Gamma} dv \partial_1 \ln K(v,v+ \Gamma ,p) )
 \\
&& \tilde{E}^-_{\Gamma} (0,F,p) = \tilde{E}^+_{\Gamma} (F,0,p) =
 \exp( + \int_{0}^{F} dv \partial_1 \ln K(v ,v + \Gamma,p) )
\label{resblocbarrier}
\end{eqnarray}

\subsection{Full measure using barriers for free boundary conditions}

The normalization (\ref{normaztoy}) concerns fixed boundary conditions
 $(u_L,u_R)$
and depends only on the difference $(u_R-u_L)$ since there is an invariance
with respect to a global shift of all the energies $u$.
For free boundary conditions, we need to sum over the end-potential $u_R$
with fixed $u_L$, which of course yields a normalization to $1$
\begin{eqnarray}
Z_{free}(x_L; x_R) = \int_{-\infty}^{+\infty} du_R Z(u_L, x_L;u_R, x_R)=1
\label{normaztoyfree}
\end{eqnarray}

The decomposition into blocks at scale $\Gamma$ now reads using 
the formulation in terms of barriers (\ref{decompofixed})
that we now denote by $F_L=u_L-u_1$, $F_1=u_2-u_1$, ... 
\begin{eqnarray}
Z_{free}(x_L; x_R)  =1  = && 
\sum_{n=1}^{+\infty} \int_0^{+\infty} dF_L \int_0^{+\infty} dF_R
\int_{\Gamma}^{+\infty}  \prod_{i=1}^{2n-2} dF_i \int_{x_L<x_1<...x_{2n-1}<x_R} \prod_{j=1}^{2n-1} dx_j  \nonumber \\
&& [ e^{ \frac{\mu^2}{12} x_L^3 -\frac{\mu}{2} x_L F_L } 
\tilde{E}^+_{\Gamma}( F_L, 0, x_1-x_L) ] \nonumber \\
&& 
 \tilde{B}^-_{\Gamma}( 0,  F_1 , x_2-x_1)
e^{ \frac{\mu}{2} x_{2} (F_1-F_2) } 
\tilde{B}^+_{\Gamma}( F_2, 0, x_3-x_{2})  \nonumber \\
&& 
 \tilde{B}^-_{\Gamma}( 0,  F_3, x_4-x_3)
e^{ \frac{\mu}{2} x_{4} (F_3-F_4) } 
\tilde{B}^+_{\Gamma}( F_{4}, 0, x_5-x_{4})
 \ldots  \nonumber \\
&& \tilde{B}^-_{\Gamma}( 0,  F_{2n-3}, x_{2n-2}-x_{2n-3})
e^{ \frac{\mu}{2} x_{2n-2} (F_{2n-3}-F_{2n-2}) } 
\tilde{B}^+_{\Gamma}( F_{2n-2}, 0, x_{2n-1}-x_{2n-2})  \nonumber \\
&& [ e^{ + \frac{\mu}{2} x_R F_R - \frac{\mu^2}{12} x_R^3 }
\tilde{E}^-_{\Gamma}( 0,  F_R, x_R-x_{2n-1}) ]
\label{decompofree}
\end{eqnarray}


\subsection{Full measure using barriers on the infinite line}

Of particular interest is of course
the limit of the infinite system $ x_L \to -\infty$ and $x_R \to \infty$.
We define the following function that describes the right boundary
in the limit $x_R \to \infty$
\begin{eqnarray}
g_{\Gamma}(x)  \equiv \lim_{x_R \to \infty} \int_{0}^{+\infty} dF_R
 [ e^{ + \frac{\mu}{2} x_R F_R - \frac{\mu^2}{12} x_R^3 }
\tilde{E}^-_{\Gamma}( 0,  F_R, x_R-x) ] 
\label{defglimit}
\end{eqnarray}
Using (\ref{resblocktoy})
\begin{eqnarray}
g_{\Gamma}(x)  =
\int_{-\infty} ^{+\infty} \frac{d\lambda}{2 \pi}
e^{- i \lambda x}
 \lim_{x_R \to \infty} 
 \left[ e^{  - \frac{\mu^2}{12} x_R^3 + i \lambda x_R}
\int_{0}^{+\infty} dF e^{ + \frac{\mu}{2} x_R F}
e^{  - \int_{0}^{F} du \partial_1 \ln K(u, u + \Gamma,i\lambda) du } \right]
\end{eqnarray}
In the limit $x_R \to \infty$, the integral over $F$ is dominated by a saddle
with $F$ large. Using (\ref{separation}) it is convenient to rewrite 
\begin{eqnarray}
\partial_1 \ln K(F,F+\Gamma,i \lambda) && =
 \partial_F \ln Ai(a F + b i \lambda ) -\psi_{\Gamma}(F,\lambda) \\
\psi_{\Gamma}(F,\lambda) && =  
\frac{ a Ai(a (F+\Gamma) + b i \lambda) }
{\pi Ai(a F + b i \lambda) 
\left[ Ai(a F + b i \lambda) Bi(a (F+\Gamma) + b i \lambda)
 -Bi(a F + b i \lambda) 
Ai(a (F+\Gamma) + b i \lambda) \right] }
\label{defpsi}
\end{eqnarray}
A saddle-point analysis with the asymptotic behavior
 (\ref{asympAi}) immediately gives
\begin{eqnarray}
\int_0^{+\infty} d\zeta e^{\zeta y} Ai(\zeta + b) 
\opsimeq_{y \to \infty} e^{\frac{1}{3} y^3 - b y} 
\end{eqnarray}
As a consequence, we get
\begin{eqnarray}
g_{\Gamma}(x)  =
\int_{-\infty} ^{+\infty} \frac{d\lambda}{2 \pi}
e^{- i \lambda x}
\frac{1}{a Ai(i b \lambda) } e^{- \int_{0}^{+\infty}
du \psi_{\Gamma}(u,\lambda)}
\label{resggamma}
\end{eqnarray}

Similarly, the other boundary $x_L \to -\infty$ yields
\begin{eqnarray}
 \lim_{x_L \to - \infty} \int_{0}^{+\infty} dF_L
[ e^{ \frac{\mu^2}{12} x_L^3 -\frac{\mu}{2} x_L F_L } 
\tilde{E}^+_{\Gamma}( F_L, 0, x-x_L) ]
= g_{\Gamma}(-x)  
\end{eqnarray}

As a consequence, the full measure using barriers on the infinite
line reads (\ref{decompofree})
\begin{eqnarray}
Z_{free}(-\infty; +\infty)  =1  = && 
\sum_{n=1}^{+\infty} 
\int_{\Gamma}^{+\infty}  \prod_{i=1}^{2n-2} dF_i \int_{x_1<...x_{2n-1}}
 \prod_{j=1}^{2n-1} dx_j  g_{\Gamma}(-x_1)  \nonumber \\
&& 
 \tilde{B}^-_{\Gamma}( 0,  F_1 , x_2-x_1)
e^{ \frac{\mu}{2} x_{2} (F_1-F_2) } 
\tilde{B}^+_{\Gamma}( F_2, 0, x_3-x_{2})  \ldots \nonumber  \\
&& \tilde{B}^-_{\Gamma}( 0,  F_{2n-3}, x_{2n-2}-x_{2n-3})
e^{ \frac{\mu}{2} x_{2n-2} (F_{2n-3}-F_{2n-2}) } 
\tilde{B}^+_{\Gamma}( F_{2n-2}, 0, x_{2n-1}-x_{2n-2}) \nonumber \\
&& g_{\Gamma}(x_{2n-1}) 
\label{decompoinfinite}
\end{eqnarray}

\section{Boltzmann equilibrium at low temperatures of the toy model} 

\label{equilibriumtoy}

In this section, we use the explicit measure of
 the renormalized landscapes of the toy model
derived in the previous section to compute various
observables that characterize the equilibrium
of the toy model (\ref{deftoy}) on the full line $]-\infty,+\infty[$
at low temperature.
For $T=0$, we need to compute the distribution over samples
of the position of absolute minimum energy on the full line.
For small $T$, we need to compute the probabilities
of configurations with nearly degenerate minima which govern 
the thermal fluctuations.

\subsection{Distribution of the position at zero temperature} 

To obtain the distribution of the position of absolute minimum energy,
we need the final state of the renormalized landscape,
corresponding to the $\Gamma \to \infty$ limit,
where all internal barriers have been eliminated.
The decomposition (\ref{decompoinfinite}) then reduces to the term
$n=1$ that contains only the boundary blocks
\begin{eqnarray}
Z_{free}(-\infty; +\infty)  =1 =  \int_{-\infty}^{+\infty} dx  
 g_{\infty}(-x) g_{\infty}(x)
\end{eqnarray}
where (\ref{resggamma})
\begin{eqnarray}
 g_{\infty}(x) = \int_{-\infty}^{+\infty} \frac{d\lambda}{2 \pi} 
\frac{e^{ - i \lambda x }}{a Ai(b i \lambda)}
\label{defg}
\end{eqnarray}
since the function $\psi_{\Gamma}$ (\ref{defpsi}) vanishes for $\Gamma \to \infty$.
The distribution of the position $x$ of the minimum 
 on $]-\infty,+\infty[$ reads
\begin{eqnarray}
&& P_{]-\infty,+\infty[}(x) = g_{\infty}(x) g_{\infty}(-x) 
\label{distriminabs}
\end{eqnarray}
in agreement with \cite{groeneboom,frachebourgmartin}.
The normalization is easily checked with the relation $b=1/a^2$
and the properties (\ref{integai1}) (\ref{asympAiimag})
\begin{eqnarray}
\int_{-\infty}^{+\infty} dx P(x) = \int_{-\infty}^{+\infty} \frac{d\lambda}{2 \pi} \frac{1}{a^2 Ai(i b  \lambda)^2}
= \frac{1}{ 2 i  }  
[\frac{Bi(z)}{Ai(z)}]^{z=+ i \infty}_{z=- i \infty}  = 1
\label{pnorma}
\end{eqnarray}
Let us now describe the asymptotic behaviors of this distribution.
The Airy function $Ai(z)$ has an infinity of poles which all are on the negative real axis :
the largest one is $e_1=- 2.33810741..$ where $Ai'(-|e_1|)=0.70121082$, 
the second one is $e_2 = - 4.08794944..$ where $Ai'(-|e_2|)=-0.80311137$ and so on.
For $x<0$, we may deform the contour in the complex plane
to express the function $g$ 
as a series over the zeroes $(- \vert e_s \vert)$ of the Airy function
\begin{eqnarray}
g_{\infty}(x <0) = a \sum_{s=1}^{+\infty} \frac{1}{Ai'(-|e_s|)} e^{- a^2 |e_s|  \vert x \vert }
\label{gmoinsinfty}
\end{eqnarray}
Thus $g(x)$ decays exponentially for $x \to -\infty$.
For $x>0$, we may shift the contour integration $\lambda \to  \lambda -i c$
\begin{eqnarray}
 g_{\infty}(x) = \int_{-\infty}^{+\infty} \frac{d \lambda}{2 \pi} 
\frac{e^{ - i \lambda x - c x }}{a Ai(b i \lambda + b c)}
\end{eqnarray}
Assuming $c$ large, we replace the function $Ai$ by its asymptotic behavior (\ref{asympAi})
\begin{eqnarray}
 g_{\infty}(x) = \frac{1}{a {\sqrt \pi} } \int_{-\infty}^{+\infty} d \lambda 
(b (c+i \lambda ))^{1/4}
e^{ -  x (c+i \lambda) + \frac{2}{3} (b (c+i \lambda))^{3/2} }
\end{eqnarray}
A saddle-point analysis leads to the choice $c= \frac{x^2}{b^3}$
and to the asymptotic behavior
\begin{eqnarray}
g_{\infty}(x ) \opsimeq_{x \to +\infty} 2 a^3 x e^{- \frac{1}{3} (a^2 x)^3 }
\label{gplusinfty}
\end{eqnarray}
Putting (\ref{gmoinsinfty}) and (\ref{gplusinfty}) together, 
the asymptotic behavior of the distribution (\ref{distriminabs}) is given by
\begin{eqnarray}
 P_{]-\infty,+\infty[}(x) \opsimeq_{ \vert x \vert \to +\infty}
 2 a^4 \vert x \vert  \frac{1}{Ai'(-|a_1|)} e^{- \frac{1}{3} (a^2 \vert x \vert)^3
- a^2 |a_1|  \vert x \vert }
\end{eqnarray}

\subsection{Moments of the position at zero temperature}

Here we obtain the explicit analytical expressions for the 
disorder averaged thermal moments moments 
 $\overline{<x^{p}>}$ of the position
 in the limit of zero temperature, where they
are determined by the absolute minimum. They can be derived from the Fourier transform
\begin{eqnarray}
\int_{-\infty}^{+\infty} dx e^{i s x} P(x) = \int_{-\infty}^{+\infty} \frac{d\lambda}{2 \pi} 
\frac{1}{a^2 Ai(i b  \lambda) Ai(i b  (\lambda+s))}
\end{eqnarray}
as (\ref{toyscale})
\begin{eqnarray}
\overline{<x^{p}>}  = b^{p} \int_{-\infty}^{+\infty} \frac{du}{2 \pi} \frac{1}{Ai( i u)}
\partial^p_{(i u)} \left(\frac{1}{Ai( i u)}\right)
 = {\mu}^{-2p/3} C_{p}
\end{eqnarray}
where $C_{p}$ are numerical constants given by 
\begin{eqnarray}
C_{p} = 2^{2p/3} \int_{-\infty}^{+\infty} \frac{du}{2 \pi} \frac{1}{Ai( i u)}
\partial^p_{(i u)} \left(\frac{1}{Ai( i u)}\right)
\end{eqnarray}
For odd $p$, an integration by parts shows that $C_{p}$ vanishes,
i.e. the odd moments vanish as it should from the symmetry $(x,-x)$.
For even $p$, we may also write
\begin{eqnarray}
C_{2p} = 2^{4p/3} \int_{-\infty}^{+\infty} \frac{du}{2 \pi} \frac{1}{Ai( i u)}
\partial^{2p}_{(i u)} \left(\frac{1}{Ai( i u)}\right)
= 2^{4p/3} \int_{-\infty}^{+\infty} \frac{du}{2 \pi} 
\left( \partial^{p}_{u} \left(\frac{1}{Ai( i u)}\right) \right)^2
\end{eqnarray}
We give the lowest moments:
\begin{eqnarray}
&& C_2 = - 2^{4/3} \int_{-\infty}^{+\infty} \frac{du}{2 \pi}
\frac{Ai'(i u)^2}{Ai(i u)^4} =
\frac{2^{4/3}}{3}  \int_{-\infty}^{+\infty} \frac{du}{2 \pi}
\frac{- i u}{Ai(i u)^2} =  1.05423856519.. \\
&& C_4 = 2^{8/3} \int_{-\infty}^{+\infty} \frac{du}{2 \pi} 
\frac{(i u - 2 \frac{Ai'(i u)^2}{Ai(i u)^2} )^2}{Ai(i u)^2}  = 3.15451..
\end{eqnarray}

These exact results are in reasonable agreement with numerical simulations
of Ref. \cite{dotsenko_mezard} which found $C_2 = 1.05381 \pm .01$ 
and $3.15441 \pm .04$. It should also be compared with the 
result of approximate replica methods which give respectively 
$C_2 = 3/(2 \sqrt{\pi})^{2/3} = 1.29038$ (RSB gaussian variational method),
$C_2 = 1.00181$ (vector breaking) \cite{dotsenko_mezard}.
Note also the exact result for the ratio $R = C_4/C_2^2=2.83827..$
in good agreement with the numerical simulation
result $R= 2.84048$, while the replica approximations
give respectively $R= 3$ (GVM) and 
$R = 2.995$ (VB).

\subsection{Joint distribution of position and energy at zero temperature}

To obtain information on the minimal energy on the full line, 
we need to fix the absolute value of the energy at a given 
point for all samples. The usual natural choice is of course
\begin{eqnarray}
U(x=0)=0
\end{eqnarray}
Then in the renormalization procedure, we need to keep information on this
value, that doesn't a priori appear in the list of extrema. 
As a consequence, to obtain the joint distribution 
$ P_{]-\infty,+\infty[}(x_0,u_0)$ of the minimal energy $U(x_0)=-u_0 
$ (with $u_0 \in [0,+\infty[$ and of its position $x_0$, 
we need to consider the renormalized
landscapes on the two independent half-lines $]-\infty,0]$ and $[0,+\infty[$
which have the same statistical properties. 
More precisely, we have for the case $x_0>0$
\begin{eqnarray}
P_{]-\infty,+\infty[}(x_0>0,u_0) = 
 P_{[0,\infty[}(x_0,u_0)
\int_0^{u_0} du_1 \int_0^{+\infty} dx_1  P_{[0,\infty[}(x_1,u_1) 
\label{twohalf}
\end{eqnarray}
where $P_{[0,\infty[}(x_0,u_0)$ is the joint probability that $(x_0,u_0)$
characterizes the absolute minimum on the half-line, and where the integrals
represents the probability that the minimum on the other half line
is less deep than $u_0$.
The decomposition (\ref{decompofree}) for the case $x_L=0$
and $x_R \to \infty$ becomes in the limit $\Gamma \to \infty$
using (\ref{defglimit})
 \begin{eqnarray}
Z_{free}(0;+\infty)  =1  =  
 \int_0^{+\infty} dF_L  \int_0^{+\infty}  dx 
 \tilde{E}^+_{+\infty}( F_L, 0, x)  g_{\infty} (x)
\label{zhalf}
\end{eqnarray}
where $g_{\infty}$ has been defined in (\ref{defg}) and where
the edge block $\tilde{E}^+$ has for Laplace transform in the $\Gamma \to \infty$ 
 \begin{eqnarray}
 \tilde{E}^+_{\infty}( F, 0, p)  = \frac{Ai( a F+ b p)}{Ai(  b p)} 
\end{eqnarray}
As a consequence, the joint distribution for the half line reads
 \begin{eqnarray}
P_{[0,\infty[}(x,u) && = \tilde{E}^+_{+\infty}( u, 0, x)  g_{\infty} (x)  
 = \int_{-\infty}^{+\infty} \frac{d\lambda_1}{2 \pi} 
\frac{e^{  i \lambda_1 x }  Ai(a u+b i \lambda_1) }{ Ai(b i \lambda_1)}
 \int_{-\infty}^{+\infty} \frac{d\lambda_2}{2 \pi} 
\frac{e^{ - i \lambda_2 x }}{a Ai(b i \lambda_2)}
\end{eqnarray}
in agreement with \cite{groeneboom}. In particular, for $u=0$
we have
 \begin{eqnarray}
P_{[0,\infty[}(x,0) && = \delta(x) g_{\infty} (0)
\label{ueq0}
\end{eqnarray}
as it should in terms of the function $g_{\infty}$ (\ref{defg}).

For the distribution of the energy $u_0$ alone, 
the relation (\ref{twohalf}) becomes after taking into account the two sides
$x>0$ and $x<0$
\begin{eqnarray}
P_{]-\infty,+\infty[}(u_0) = 
2 P_{[0,\infty[}(u_0) \int_0^{u_0} du   P_{[0,\infty[}(u) 
= \frac{d}{du_0} \left[ \int_0^{u_0} du   P_{[0,\infty[}(u) \right]^2
\label{twohalfbis}
\end{eqnarray}
where
 \begin{eqnarray}
P_{[0,\infty[}(u) && =  \int_0^{+\infty}  dx P_{[0,\infty[}(x,u) 
 = \lim_{p \to 0} 
\left(  \int_{-i\infty}^{+i\infty} \frac{dz_2}{2 i \pi} 
\frac{ 1 }{a Ai(b z_2)}
 \int_{-i\infty}^{+i\infty} \frac{dz_1}{2 i \pi} 
\frac{  Ai(a u+b z_1) }{ Ai(b z_1)}
\frac{1}{ p + z_2-z_1 }  \right)
\end{eqnarray}
is the distribution of the energy alone on the half-line $[0,+\infty[$.
For $u>0$, we may close the contour in $z_1$ on the right at infinity
to obtain via Cauchy theorem
 \begin{eqnarray}
P_{[0,\infty[}(u) &&=  \int_{-i\infty}^{+i\infty} \frac{dz_2}{2 i \pi} 
\frac{  Ai(a u+b z_2) }{a Ai^2(b z_2)}
= \int_{-\infty}^{+\infty} \frac{d \lambda}{2 \pi} 
\frac{  Ai(a u+b i \lambda) }{a Ai^2(b i \lambda)}
\end{eqnarray}
and this expression is also valid for $u=0$ (\ref{ueq0}).

The expansion near $u=0$ reads
 \begin{eqnarray}
P_{[0,\infty[}(u) && \opsimeq_{u \to 0} 
 \int_{-\infty}^{+\infty} \frac{d \lambda}{2 \pi} 
\frac{  Ai(b i \lambda)+ a u Ai'(b i \lambda)+ \frac{a^2 u^2}{2} Ai''(b i \lambda) }{a Ai^2(b i \lambda)}
 = g_{\infty} (0) - \frac{ u^2}{2  } 
 g_{\infty}' (0) +O(u^3)
\end{eqnarray}
which leads to the following expansion for the full line
\begin{eqnarray}
P_{]-\infty,+\infty[}(u_0) \opsimeq_{u_0 \to 0} 
2 u_0 g_{\infty}^2 (0) + O(u_0^3)
\end{eqnarray}

To obtain the asymptotic behavior for large $u$, we may shift the contour integration $\lambda \to  \lambda -i c$, and assuming $c$ large,
we get via the saddle-point method with the choice $c=\frac{au}{3b}$ (\ref{asympAi})
 \begin{eqnarray}
P_{[0,\infty[}(u) && =
 \int_{-\infty}^{+\infty} \frac{d \lambda}{2 \pi} 
\frac{  Ai(a u+b i \lambda+bc) }{a Ai^2(b i \lambda+bc)}
\opsimeq_{c \to \infty}
 \frac{1}{a \sqrt \pi} \int_{-\infty}^{+\infty} d \lambda 
\frac{ (b i \lambda+bc)^{1/2}}{  (a u+b i \lambda+bc)^{1/4} }
e^{ \frac{4}{3}(b i \lambda+bc)^{3/2} - \frac{2}{3}
(a u+b i \lambda+bc)^{3/2} } \\
&& \opsimeq_{u \to \infty} \frac{2}{3} a (au)^{1/2} 
e^{-4 \left( \frac{au}{3} \right)^{3/2} }
\end{eqnarray}

\subsection{ Samples presenting nearly degenerate minima
and thermal fluctuations at low temperature}

As explained in the introduction, there are remarquable identities
\cite{identities88} for the disorder averages of the thermal cumulants 
of the position (\ref{genlambda2}). It was moreover shown in \cite{identities88}
that the result (\ref{second}) for the second cumulant at low
temperature contains a finite contribution from the rare configurations
presenting two nearly degenerate minima. In this section, 
we compute the statistical properties of such configurations
and show that they actually give the full exact result (\ref{second}).

\subsubsection{Configurations with two nearly degenerate minima}

It is shown in Appendix (\ref{twodege})
that the probability to have two nearly degenerate minima separated
by an energy difference $\Delta E=\epsilon \to 0$
and situated at $x_1$ and $x_2$ reads 
\begin{eqnarray}
{\cal D}(\epsilon,x_1,x_2) =  \epsilon g_{\infty}(-x_1) d(x_2-x_1)
g_{\infty}(x_2) +O(\epsilon^2)
\label{restwomin}
\end{eqnarray}
where the function $g_{\infty}$ has been defined in (\ref{defg})
and where
\begin{eqnarray}
d(y) =a  \int_{-\infty}^{+\infty} \frac{d \lambda}{2 \pi}
 e^{i \lambda y }
 \frac{ Ai'(i b \lambda)}{ Ai(i b\lambda)}
\end{eqnarray}

The total probability density 
to have two degenerate minima separated by a distance $y>0$ thus reads
\begin{eqnarray}
D(y) && = \int_{-\infty}^{+\infty} dx_1 
\lim_{\epsilon \to 0} \left( \frac{ {\cal D}(\epsilon,x_1,x_1+y)}{\epsilon}
\right)   = d(y) \int_{-\infty}^{+\infty} dx_1 
g(-x_1)  g (x_1+y)  = b d(y) \int_{-\infty}^{+\infty} \frac{d\lambda}{2 \pi}
\frac{e^{-i \lambda y} }{  Ai^2( i b \lambda)} 
\end{eqnarray}

In Laplace with respect to $y$ we have
\begin{eqnarray}
{\hat  D}(p) && \equiv \int_0^{+\infty} dy e^{-p y}  D(y)
 = a  
\int_{- i \infty}^{+ i \infty} \frac{d z_2}{2 \pi i}
\frac{1 }{  Ai^2(z_2)}
  \int_{- i \infty}^{+i \infty} \frac{d z_1}{2 \pi i }
 \frac{ Ai'(z_1)}{ Ai(z_1)}
\ \ \frac{1}{b p + z_2 - z_1}
\end{eqnarray}

In particular, the second moment is given by
\begin{eqnarray}
\int_0^{+\infty} dy y^2 D(y) 
&& = \lim_{p \to 0} \left( 
\frac{ d^2 {\hat D}(p)}{dp^2 } \right) \\
&& = \lim_{p \to 0} \left( 
 a  
\int_{- i \infty}^{+ i \infty} \frac{d z_2}{2 \pi i}
\frac{1 }{  Ai^2(z_2)}
  \int_{- i \infty}^{+i \infty} \frac{d z_1}{2 \pi i }
 \frac{ Ai'(z_1)}{ Ai(z_1)}
\ \ \frac{ 2 b^2 }{( b p + z_2 - z_1 )^3}   \right)  \\
&& =  
 a b^2 \int_{- i \infty}^{+ i \infty} \frac{d z_2}{2 \pi i}
\frac{1 }{  Ai^2(z_2)}
 \left(  \frac{ Ai'}{ Ai} \right)''(z_2)
\label{deri2}
\end{eqnarray}
where we have used Cauchy theorem. This last integral may be computed via integration by part, with the help of the following properties of
Airy function :
the function $f(z) = \frac{Ai'(z)}{Ai(z)}$ satisfies $f'(z)=z-f^2(z)$
and the function $g(z) =\frac{1}{Ai^2(z)}$ satisfies $g'(z)=-2 f(z) g(z)$ :
\begin{eqnarray}
\int_0^{+\infty} dy y^2 D(y) 
&& =  
\frac{ a b^2 }{2} \int_{- i \infty}^{+ i \infty} \frac{d z_2}{2 \pi i}
\frac{1 }{  Ai^2(z_2)} = \frac{ a b^2 }{2} 
\left[ \frac{ Bi(z) }{2i Ai(z) } \  \right]_{ z=- i \infty}^{z=+ i \infty} 
 = \frac{ a b^2 }{2} = \frac{1}{\mu}
\label{y2toy}
\end{eqnarray}
with (\ref{toyscale}).

\subsubsection{Contribution to the second thermal cumulant at low temperature}

The contribution of these configurations with two degenerate minima
to the second thermal cumulant of the position (\ref{second})
can be estimated at order $T$ in temperature as follows.
The two minima have for respective Boltzmann weights 
$ p =\frac{ 1 } { 1 + e^{ - \beta \epsilon} }$ and
 $(1-p)=\frac{  e^{ - \beta \epsilon} } { 1 + e^{ - \beta \epsilon} }$.
The variable $(x-<x>)$ is $(1-p)(x_1-x_2)$ with probability $p$ and $p (x_2-x_1)$ with probability $1-p$, so we get after averaging 
over $\epsilon$ with a flat measure as in 
the study of the rare events in \cite{us_long,us_long_rf}

\begin{eqnarray}
\overline{ < (x-<x>)^2 > } && = \overline{ p(1-p) (x_1-x_2)^2 } \\
&&  = \int_{-\infty}^{+\infty} dx_1 \int_{x_1}^{+\infty} dx_2  
\int_{-\infty}^{+\infty} d\epsilon {\cal D}'(\epsilon=0,x_1,x_2) \frac{e^{- \epsilon/T}}{(1 + e^{- \epsilon/T})^2}   (x_2-x_1)^2  \\
&& = T \int_0^{+\infty} dy   D(y)  y^2  
\label{two}
\end{eqnarray}
Using (\ref{y2toy}), we exactly find the exact result (\ref{second}).
This shows that the thermal fluctuations at low temperature
are entirely due the presence of metastable states 
in rare disordered samples. In particular, the susceptibility
\begin{eqnarray}
\chi \equiv \frac{1}{T} \left( <x^2>-<x>^2 \right)
\label{suscepti}
\end{eqnarray}
has for average a finite value at zero-temperature
\begin{eqnarray}
\overline{ \chi } = \frac{1}{\mu}
\end{eqnarray}
but only the samples with two nearly degenerate minima contribute
to this zero-temperature value. The typical samples with only
one minimum have a susceptibility that vanishes at zero temperature.
This type of behavior was also found numerically in \cite{mezardcorresdp}
and analyzed in \cite{hwa} for the
 random directed polymer model. 

\subsubsection{Contribution to the thermal moments at low temperature}

The small-T behavior
 of the even moments of $(x-<x>)$ can be similarly computed
\begin{eqnarray}
 \overline {  < (x-<x>)^{2n} > } && = \frac{T}{n} \int_0^{+\infty} y^{2n} D(y) +O(T^2)
 \label{momentslowT}
\end{eqnarray}
The moments of $D(y)$ read, with the same method as in 
(\ref{deri2}) 
\begin{eqnarray}
\int_0^{+\infty} dy y^k D(y) 
&& =  
 a (-1)^k b^k \int_{- i \infty}^{+ i \infty} \frac{d z}{2 \pi i}
\frac{1 }{  Ai^2(z)}
 \left(  \frac{ Ai'}{ Ai} \right)^{(k)}(z)  
 = \frac{ \sigma^{(k-2)/3} }{ \mu^{(2k-1)/3} } m_k
\label{derik}
\end{eqnarray}
where $m_k$ are numbers given by
\begin{eqnarray}
m_k =  (-1)^k 2^{(2k-1)/3}  \int_{- i \infty}^{+ i \infty} \frac{d z}{2 \pi i}
\frac{1 }{  Ai^2(z)}
 \left(  \frac{ Ai'}{ Ai} \right)^{(k)}(z)  
\end{eqnarray}

\subsubsection{Discussion}

The comparison between (\ref{momentslowT}) and (\ref{genlambda2}) 
shows that there are 
a lot of cancelations in the disorder averages of the cumulants.
This phenomenon was already emphasized in \cite{huse}
for the random directed polymer, and will be studied in more
details in \cite{us_identities} for the toy model as well
as in other one-dimensional random systems which do not
present the statistical tilt symmetry leading to the identities (\ref{genlambda2}).   

Another question that needs further explanations and that will be
studied in \cite{us_identities} is why the result (\ref{second})
which can be understood at order $T$ by considering only the configurations
with two degenerate minima remains true for arbitrary temperature.
Indeed, in the light of the above discussion, it would be natural to expect
corrections at order $T^2$ coming from configurations
with three degenerate minima and described by the
distribution function analogous to (\ref{restwomin})
\begin{eqnarray}
{\cal T}(\epsilon_1,\epsilon_2,x_1,x_2,x_3) =  \epsilon_1 \epsilon_2 g_{\infty}(-x_1) d(x_2-x_1) d(x_3-x_2)
g_{\infty}(x_3) 
\end{eqnarray}
and so on for configurations with $n$ nearly degenerate minima.
One also expect corrections from the thermal width around minima.
The study of the roles played by these various effects 
in different systems is postponed to a future work \cite{us_identities}.

\section{Non-equilibrium dynamics of the toy model}

\label{dynamicstoy}

In this section, we study the non-equilibrium dynamics of the toy model
by using the full measure of the
renormalized landscape of the toy model for an arbitrary scale $\Gamma$.
We also briefly summarize some results for the RFIM in a field gradient,
already discussed in the Introduction. 

As in the Sinai model \cite{us_long}, the long-time dynamics is 
asymptotically entirely determined by the renormalized landscape
at scale
\begin{eqnarray}
\Gamma = T \ln t
\label{correst}
\end{eqnarray}
in the sense that a particle starting at $x_0$ at time $t=0$
will be at time $t$ around the minimum of the renormalized valley
at scale $\Gamma$ that contains the initial position $x_0$.
As in the Sinai model \cite{us_long}, this method allows to compute
asymptotically exact results for scaling variables
in the limit $t \to \infty$
corresponding to large barriers, and to large distances.
Here in the toy model, there is a typical scale of energy given by $(1/a)$
and a typical length $b=1/a^2$ (\ref{toyscale}).
As a consequence, we need to define the following scaling variables
for the barrier
\begin{eqnarray}
\gamma \equiv a \Gamma = ( \frac{\mu}{2} )^{1/3} T \ln t  
\label{rescalegamma} 
\end{eqnarray}
and for the length
\begin{eqnarray}
 \quad \tilde{x} = \frac{ x }{b} =  ( \frac{\mu}{2} )^{2/3} x 
\label{rescalx}
\end{eqnarray}
The RSRG method will thus give asymptotic exact results
for the dynamics in the double limit $t \to \infty$ 
(to have large barriers $\Gamma$) and $\mu \to 0$ 
(to have large typical scales for energy and length in the toy landscape)
 with the scaling variables $\gamma$ and $\tilde{x}$ fixed.
For large $\gamma$, as studied in previous section,
the system will reach equilibrium. For small $\gamma$, i.e.
 when the barrier $\Gamma=T \ln t$
is much smaller than the characteristic energy scale $(1/a)$ 
of the toy model, the dynamics 
should look like the Sinai model, where the typical length
associated to the dynamics is $\Gamma^2$ \cite{us_long} 
so that the scaling variable for distance 
of diffusion during the time $t$ reads in this regime
\begin{eqnarray}
 (\Delta X)_{Sinai} = \frac{\Delta x} {\Gamma^2}
 = \frac{ \Delta \tilde x }{\gamma^2}  
\label{rescalsinai}
\end{eqnarray}
We will first consider a global quantity of the whole system :
 each sample is characterized by
an equilibration time $t_{eq}$ corresponding 
to the largest barrier $\Gamma_{eq}$ in the system via
\begin{eqnarray}
\Gamma_{eq} = T \ln t_{eq}
\end{eqnarray}
We will compute the distribution over samples of this equilibration time
and then characterize the samples which have an anomalously large
equilibration time and which thus dominate the long time dynamics.
We will then study local quantities, such as the probability density of minima of the renormalized landscape, and the diffusion front, i.e. the generalization of the 
Kesten distribution of the Sinai model \cite{kesten,us_long}
to the case of the toy model.

As discussed in the Introduction, the equilibrium state for the RFIM in a field
gradient for large $J$ is obtained by stopping the RSRG 
at scale $\Gamma = \Gamma_{eq0} = T \ln t_{eq0} = 2 J$
at which equilibrium is attained and putting a $B$ domain 
wall in each minimum and a $A$ domain in each maximum 
of the renormalized landscape at scale $\Gamma = 2 J$. The equilibrium state
thus consists in a $B$ domain plus $n$ additional pairs of domains $A-B$,
the probability of $n$ with respect to sample being thus
 $N^{(2+2 n)}_{\Gamma=2 J}$.
The distribution of the total number of domains at equilibrium 
is thus a scaling function of
$\gamma_{eq0} = a 2 J$. Below we obtain the behavior
 at large $\gamma_{eq0}$ of these probabilities for
$n=0,1$ as well as an estimate of the total number of domains at
small $\gamma_{eq0}$. Similarly, inserting instead $\Gamma = T \ln t$,
 one also finds these probabilities for the out of equilibrium system
 at time $t$ starting
from a random (i.e. high temperature) configuration at time zero. The total magnetization reads
\begin{eqnarray}
M = 2 \sum_{\alpha=1}^{N_A} a_\alpha - 2 \sum_{\alpha=1}^{N_B} b_\alpha
\end{eqnarray}
its probability measure at time $t$ can thus be formally expressed using the full measure for the renormalized landscape at $\Gamma = T \ln t$, and the equilibrium expression
obtained using $\Gamma = 2J$. In the limit of large $\gamma_{eq0}$, its probability 
distribution coincides exactly with the one of the absolute minimum of the toy
model given by formula (\ref{distriminabs}) setting $M = 2 x$.

For non random initial configurations prepared with a single domain $B$, the
equilibration time of this domain is $t_{eq}$ computed below (provided 
$T \ln t_{eq0} > T \ln t_{eq}$, otherwise additional pairs are created at
time $t_{eq0}$ and take positions in the additional deep extrema). If the
single domain is $A$, then it is expelled from the system when the extrema corresponding 
to the highest barrier is decimated, again the expulsion time is the same
$T \ln t_{exp} = T \ln t_{eq}$. Note, interestingly, that if two almost degenerate minima (within
energy difference $T$) exist, the domain wall $A$ will be expelled
with finite probabilities towards the two boundaries,
and the system will be
 in a macroscopic mixture of two reversed states
(either all spins (+) or all (-)) between times $t_{exp}$ and $t_{eq0}$ (if we assume that $h L \ll J$).

\subsection{Distribution over samples of the equilibration time}

We first study the distribution over samples of the equilibration time $t_{eq}$
at which equilibrium is reached.
 The probability ${\cal P}(t_{eq}<t)$
that the system has already reached equilibrium at 
time $t$ is directly given by the probability
 that there remains only one valley at renormalization
scale $\Gamma = T \ln t$ in the decomposition (\ref{decompoinfinite})
with (\ref{resggamma})
\begin{eqnarray}
{\cal P}(t_{eq}<t) && = 
\int_{-\infty}^{+\infty} dx g_{\Gamma}(-x) g_{\Gamma}(x) = \int_{-\infty}^{+\infty} \frac{d\lambda}{2 \pi}
\frac{1}{a^2 Ai^2(i b \lambda) } e^{- 2 \int_{0}^{+\infty}
dF \psi_{\Gamma}(F,\lambda)}
\label{ptequi1}
\end{eqnarray}
where $\psi_{\Gamma}$ has been defined in (\ref{defpsi}).

It is convenient to rewrite it as a scaling function of 
$\gamma$ defined in (\ref{rescalegamma})
\begin{eqnarray}
{\cal P}(t_{eq}<t)  = \Phi (\gamma)
  = \int_{-\infty}^{+\infty} \frac{d\lambda}{2 \pi}
\frac{1}{ Ai^2(i  \lambda) } e^{- 2 \int_{0}^{+\infty}
 df \tilde{\psi}_{\gamma} (f,\lambda)} 
\label{ptequi}
\end{eqnarray}
with 
\begin{eqnarray}
 \tilde{\psi}_{\gamma}(f,\lambda) = \frac{  Ai(f+\gamma +  i \lambda) }
{\pi Ai(f +  i \lambda) 
\left[ Ai(f +  i \lambda) Bi(f+\gamma +  i \lambda)
 -Bi(f +  i \lambda) 
Ai(f+\gamma +  i \lambda) \right] }
\label{defrescalpsi}
\end{eqnarray}

Equivalently, this result describes the probability distribution $P_{max}(\Gamma)$ over samples of
the largest barrier $\Gamma=T \ln t_{eq}$
existing in the system
\begin{eqnarray}
P_{max}(\Gamma) = \frac{d}{d \Gamma} {\cal P}(t_{eq}<t) =
 \frac{d}{d \Gamma} \Phi \left( a \Gamma \right)
\label{distribarrmax}
\end{eqnarray}

\subsubsection{Probability of large equilibration times}

For large $ \gamma$, the dominant behavior
of the function $\tilde{\psi}_{\gamma}$ (\ref{defrescalpsi})
is found by using (\ref{asympAi})
\begin{eqnarray}
 \frac{Ai ( \gamma+f +  i \lambda )}
{Bi ( \gamma+f +  i \lambda )}
\opsimeq_{\gamma \to \infty} \frac{1}{2} e^{- \frac{4}{3} \left(
 \gamma+f +  i \lambda ) \right)^{3/2}}
\end{eqnarray}
which leads to
\begin{eqnarray}
\tilde{\psi}_{\gamma}(f,\lambda) && = \opsimeq_{\gamma \to \infty} 
\frac{ a e^{- \frac{4}{3} \left(
 \gamma+f +  i \lambda ) \right)^{3/2}} }
{2 \pi Ai^2(f +  i \lambda) 
\left[  1
 - e^{- \frac{4}{3} \left(
 \gamma+f +  i \lambda ) \right)^{3/2}}
\left( \frac{Bi(f +  i \lambda)}{2 Ai(f +  i \lambda)} \right)
 \right] } \\
&& \opsimeq_{\gamma \to \infty} \frac{ 1} {2 \pi Ai^2(f +  i \lambda)  } 
e^{- \frac{4}{3} \gamma^{3/2}
- 2 \gamma^{1/2}( f +  i \lambda )
 +O \left(\frac{1}{ \gamma^{1/2} } \right) }
\label{psiglarge}
\end{eqnarray}
The change of variable $f= \frac{w}{ \sqrt{ \gamma}}$
 gives
\begin{eqnarray}
\int_0^{+\infty} df {\tilde \psi}_{\gamma}(f,\lambda)
&& \opsimeq_{\gamma \to \infty}
\int_0^{+\infty} \frac{d w}{ \sqrt{ \gamma}}
 \frac{ 1} {2 \pi Ai^2(  i \lambda+ \frac{w}{ \sqrt{ \gamma}})  } 
e^{- \frac{4}{3} \gamma^{3/2}
- 2i    \lambda   \gamma^{1/2}
- 2 w } \left[1
 +O \left(\frac{1}{ \gamma^{1/2} } \right)  \right]  \\
&& \opsimeq_{\gamma \to \infty}
 \frac{1}{4 \pi   \sqrt{ \gamma} Ai^2(  i \lambda)  } 
e^{- \frac{4}{3} \gamma^{3/2}
- 2i    \lambda   \gamma^{1/2} }
\left[1
 +O \left(\frac{1}{ \gamma^{1/2} } \right)  \right] 
\end{eqnarray}

The leading behavior of (\ref{ptequi}) is thus given by
\begin{eqnarray}
 \Phi (\gamma) && \opsimeq_{\gamma \to \infty}
 \int_{-\infty}^{+\infty} \frac{d\lambda}{2 \pi}
\frac{1}{ Ai^2(i  \lambda) } 
\left[ 1 - 
  \frac{1}{2 \pi   \sqrt{ \gamma} Ai^2(  i \lambda)  } 
e^{- \frac{4}{3} \gamma^{3/2}
- 2i    \lambda  \gamma^{1/2} }
\left[1
 +O \left(\frac{1}{\gamma^{1/2} } \right)  \right] \right] \\
 && \opsimeq_{\gamma \to \infty} 1- 
 \frac{ e^{- \frac{4}{3} \gamma^{3/2} } }{2 \pi   \sqrt{ \gamma}   } 
\int_{-\infty}^{+\infty} \frac{d\lambda}{2 \pi}
\frac{e^{- 2i    \lambda  \gamma^{1/2} }}{ Ai^4(i  \lambda) }
\left[1
 +O \left(\frac{1}{\gamma^{1/2} } \right)   \right] 
\label{ptlargeequi} 
\end{eqnarray}

Shifting the contour integration $\lambda \to  \lambda -i c$
and assuming $c$ large, we get using (\ref{asympAi})
\begin{eqnarray}
\int_{-\infty}^{+\infty}
\frac{d\lambda}{2 \pi}
\frac{e^{ - i \lambda 2  \sqrt{ \gamma}} }
{  Ai^4( i \lambda) }
 = \int_{-\infty}^{+\infty} \frac{d \lambda}{2 \pi} 
\frac{e^{ -  2  \sqrt{ \gamma} (c+i \lambda) }}{ Ai^4( i \lambda +  c)}
 \opsimeq_{c \to \infty} 8 \pi \int_{-\infty}^{+\infty} d \lambda 
 (c+i \lambda )
e^{ -  2  \sqrt{ \gamma} (c+i \lambda) + \frac{8}{3}  (c+i \lambda)^{3/2} } 
\end{eqnarray}
A saddle-point analysis leads to the choice $c= \frac{ \gamma}{ 4 }$
and to the asymptotic behavior
\begin{eqnarray}
\int_{-\infty}^{+\infty}
\frac{d\lambda}{2 \pi}
\frac{e^{ - i \lambda 2  \sqrt{ \gamma}} }
{  Ai^4( i \lambda) }
\opsimeq_{\gamma \to +\infty} \pi \sqrt{2 \pi}  \gamma^{5/4} 
e^{- \frac{1}{6} \gamma^{3/2}}
\label{g4largeg}
\end{eqnarray}

The probability to have only one valley at time $t$
finally behaves for large $\gamma$ as (\ref{ptlargeequi})
\begin{eqnarray}
{\cal P}(t_{eq}<t) = \Phi(\gamma)
&& \opsimeq_{\gamma \to \infty}= 1-  \sqrt{\frac{\pi}{2}}
\gamma^{3/4} e^{- \frac{3}{2} \gamma^{3/2}}
\label{resptlargeequi}
\end{eqnarray}

Equivalently, this result describes the 
asymptotic tail of
probability distribution $P_{max}(\Gamma)$ over samples of
the largest barrier $\Gamma=T \ln t_{eq}$
existing in the system (\ref{distribarrmax})
\begin{eqnarray}
P_{max}(\Gamma)  =
 \frac{d}{d \Gamma} \Phi \left( a \Gamma \right)
&& \opsimeq_{\Gamma \to \infty} \frac{9}{4} a  \sqrt{\frac{\pi}{2}}
(a \Gamma)^{5/4} e^{- \frac{3}{2} (a \Gamma)^{3/2}}
\label{distribarrmaxlarge}
\end{eqnarray}

\subsubsection{Small equilibration times}

For small $ \gamma$, the leading behavior of (\ref{ptequi})
can be found with the change of variable
 $f= \frac{s^2}{  \gamma^2}$ using (\ref{asympAi})
\begin{eqnarray}
&&  \int_0^{+\infty} df {\tilde \psi}_{\gamma}(f,\lambda) = 
 \int_0^{+\infty} \frac{2 s ds}{  \gamma^2 }
 \frac{ Ai(\frac{s^2}{  \gamma^2}
 +  i \lambda+ \gamma) }
{\pi Ai(\frac{s^2}{  \gamma^2} +  i \lambda) 
\left[ Ai(\frac{s^2}{  \gamma^2} +  i \lambda) 
Bi(\frac{s^2}{  \gamma^2} +  i \lambda+ \gamma)
 -Bi(\frac{s^2}{  \gamma^2} +  i \lambda) 
Ai(\frac{s^2}{  \gamma^2} +  i \lambda + \gamma) \right] }\\
&& \opsimeq_{\gamma \to 0}
\frac{2}{  \gamma^3 } \int_0^{+\infty} ds s^2 \frac{e^{-s}}{\sinh s}
= \frac{\zeta(3)}{  \gamma^3 }
\end{eqnarray}
where $\zeta(n)$ is the Riemann zeta function.
This leads to the following essential singularity
for the short equilibration times
\begin{eqnarray}
{\cal P}(t_{eq}<t)  = \Phi(\gamma) \opsimeq_{\gamma \to 0}
e^{- 2 \frac{\zeta(3)}{  \gamma^3 } }
\label{ptsmallequi} 
\end{eqnarray}
or equivalently for 
probability distribution $P_{max}(\Gamma)$ of
the largest barrier $\Gamma$
existing in the system (\ref{distribarrmax})
\begin{eqnarray}
P_{max}(\Gamma)  =
 \frac{d}{d \Gamma} \Phi \left( a \Gamma \right)
&& \opsimeq_{\Gamma \to 0} \frac{6 \zeta(3)}{  a \Gamma^4 }
e^{- 2 \frac{\zeta(3)}{ ( a \Gamma)^3 } }
\label{distribarrmaxsmall}
\end{eqnarray}

\subsection{Rare configurations with anomalously large equilibration times}

We are now interested into the statistical properties of the configurations
that present an anomalously large equilibration time, 
as described by the tail (\ref{resptlargeequi}) since these configurations
are important for the long time dynamics.
In the regime $ \gamma \gg 1$ under consideration, the leading correction to 
the one valley event (\ref{resptlargeequi})
that we have computed in the previous section
corresponds to the two valleys configurations, 
because the configurations with three or more valleys are even more unlikely. 
These configurations correspond to the last non-equilibrium dynamics in the system before it finally equilibrates by going over the largest barrier $\Gamma_{eq}$. We can already deduce 
from the previous section the total probability of the two valleys configurations (\ref{resptlargeequi})
\begin{eqnarray}
N^{(4)}_{\Gamma} \opsimeq_{\Gamma \to \infty}  \sqrt{\frac{\pi}{2}}
(a \Gamma)^{3/4} e^{- \frac{3}{2} (a \Gamma)^{3/2}}
\label{n4norma}
\end{eqnarray}

To characterize the statistical properties of these two valley events,
we have the probability measure (\ref{decompoinfinite}) 
\begin{eqnarray}
N^{(4)}_{\Gamma}(x_1,F_1,x_2,F_2,x_3) 
&& = g_{\Gamma}(-x_1) 
 \tilde{B}^-_{\Gamma}( 0,  F_1 , x_2-x_1)
e^{ \frac{\mu}{2} x_{2} (F_1-F_2) } 
\tilde{B}^+_{\Gamma}( F_2, 0, x_3-x_{2})  g_{\Gamma}(x_3) 
\end{eqnarray}
where $x_1<x_2<x_3$, $x_1$ and $x_3$ are the two local minima
separated by the barriers $F_1\geq\Gamma$ and $F_2\geq\Gamma$
 from the local maximum $x_2$. 
For large $\Gamma$, the leading behavior of the blocks $\tilde{B}$ (\ref{resblocbarrier}) can be found using (\ref{kfactor})
\begin{eqnarray}
K(u,v,p) \opsimeq_{v >> u} \frac{\pi}{a}
 Ai(a u + b p) Bi(a v + b p) 
\label{asymptoktoy}
\end{eqnarray}
which yields
\begin{eqnarray}
&& \tilde{B}^-_{\Gamma} (0,F,p) = \tilde{B}^+_{\Gamma} (F,0,p) 
\opsimeq_{\Gamma \to \infty} 
 \frac{a}{\pi Ai(bp) Bi(a F+bp)} 
\label{asymptoblocb}
\end{eqnarray}
At the same level on approximation, $g_{\Gamma}$ is given by $g_{\infty}$
(\ref{defg}), so that we finally get at leading order
\begin{eqnarray}
N^{(4)}_{\Gamma}(x_1,F_1,x_2,F_2,x_3) 
&& \opsimeq_{\Gamma \to \infty} 
\frac{ e^{ \frac{\mu}{2} x_{2} (F_1-F_2) } }{\pi^2}
 \int_{-\infty}^{+\infty} \frac{d\lambda_1}{2 \pi} 
 \int_{-\infty}^{+\infty} \frac{d\lambda_2}{2 \pi} 
\int_{-\infty}^{+\infty} \frac{d\lambda_3}{2 \pi} 
\int_{-\infty}^{+\infty} \frac{d\lambda_4}{2 \pi} \\
&& \frac{e^{  i (\lambda_1+\lambda_2) x_1 +  i (\lambda_3-\lambda_2) x_2 
-i (\lambda_2+\lambda_3)x_3 }}{ Ai(b i \lambda_1) Ai(b \lambda_2)
 Bi(a F_1+b\lambda_2) Ai(b \lambda_3) Bi(a F_2+b\lambda_3) Ai(b i \lambda_4)}
\label{fulln4}
\end{eqnarray}

It is shown in Appendix (\ref{2valleys}) that
after integration over the barriers and the position $x_2$
of the local maximum, the measure for the two minima $(x_1,x_3)$ 
has the simple structure
\begin{eqnarray}
M^{(4)}_{\Gamma}(x_1,x_3) && \equiv \int_{x_1}^{x_3} dx_2  \int_{\Gamma}^{+\infty} dF_1
\int_{\Gamma}^{+\infty} dF_2 N^{(4)}_{\Gamma}(x_1,F_1,x_2,F_2,x_3) \\
&&  \opsimeq_{\Gamma \to \infty} 
\frac{  a^2  }{ 2 \pi (a \Gamma)^{1/2} }  e^{-\frac{4}{3} (a \Gamma)^{3/2} }
g_{\infty}(-x_1)  g_{\infty}^{(2)}(x_1 -x_3+ 2 b (a \Gamma)^{1/2}) g_{\infty}(x_3) 
\label{restwominvall}
\end{eqnarray}
with
\begin{eqnarray}
g_{\infty}^{(2)}(x)=
\int_{-\infty}^{+\infty}
\frac{d\lambda}{2 \pi}
\frac{e^{ - i \lambda x} }
{ a^2 Ai^2(b i \lambda) } =(g_{\infty}*g_{\infty})(x)
\end{eqnarray}
The total normalization $N^{(4)}$ (\ref{n4norma}) is recovered by
integration over $(x_1,x_3)$ with the use of (\ref{g4largeg}). 

Since all the functions $g_{\infty}(y)$ are centered around $y=0$,
the two-minima measure $M^{(4)}_{\Gamma}(x_1,x_3)$ is concentrated around
$(x_1,x_3)=(-x_{\Gamma},x_{\Gamma})$ with
\begin{eqnarray}
 x_{\Gamma} =  \frac{\sqrt{a \Gamma}}{ 2 a^2} 
\label{n4scalingmin}
\end{eqnarray}
More precisely, a saddle-point analysis leads to
the approximation by a product of two Gaussian distributions
of broadness
\begin{eqnarray}
 \sigma_{\Gamma} = \frac{\sqrt 3}{2 a^2 (a \Gamma)^{1/4} }
\label{sigma}
\end{eqnarray}
as
\begin{eqnarray}
M^{(4)}_{\Gamma}(x_1,x_3) 
&&  \opsimeq_{\Gamma \to \infty} 
 = N^{(4)}_{\Gamma}
 \frac{e^{  - \frac{1}{2 \sigma^2_{\Gamma} }  (x_1+x_{\Gamma})^2} }
{2 \sqrt {2 \pi \sigma^2_{\Gamma} }} 
 \frac{e^{  - \frac{1}{2 \sigma^2_{\Gamma} }  (x_3-x_{\Gamma})^2} }
{2 \sqrt {2 \pi \sigma^2_{\Gamma} }} 
\label{gauss}
\end{eqnarray}
concentrating all the normalization $N^{(4)}_{\Gamma}$ found above (\ref{n4norma}). 

The two valleys configurations with an anomalously
 large barrier $a \Gamma \gg 1$
thus present minima which are anomalously far from the origin 
when compared to the typical scale of minima (\ref{toyscale})
\begin{eqnarray}
x^{(4)}_{\rm{min}} =  \frac{b \sqrt{a \Gamma}}{ 2 } \gg b
\end{eqnarray}
which is necessary with the Brownian scaling to have a barrier 
of order $\Gamma$ between them.

\subsection{ Probability density of minima in the renormalized landscape}

The probability $V_{\Gamma}(x)$
that the point $x$ is the bottom of a renormalized valley at scale $\Gamma$
can be decomposed into
\begin{eqnarray}
V_{\Gamma}(x)=V_{\Gamma}^L(x) V_{\Gamma}^R(x)
\label{probavgamma}
\end{eqnarray}
where $V_{\Gamma}^R(x)$ represents the constraints
 on the landscape on the half line $[x,+\infty[$ : 
 the potential starting from $U(x)$ has to reach $U(x)+\Gamma$
at an arbitrary position $y$ without any return to $U(x)$,
which reads in terms of the Fokker-Planck block (\ref{changetotilde},\ref{defhtoy})
with the use of (\ref{globalshift},\ref{resblocbarrier},\ref{ktoy})
\begin{eqnarray}
V_{\Gamma}^R(x) && = \int_x^{+\infty} dy B_{\Gamma}^-(0,x;\Gamma,y)
 = \int_x^{+\infty} dy e^{ h(0,x)-h(\Gamma,y) } 
{ \tilde B}_{\Gamma}^-(0,\Gamma;y-x)  \\
&& = e^{ \frac{\mu^2}{12 } x^3 } 
\int_{-\infty}^{+\infty} \frac{d \lambda}{2 \pi} 
\frac{ a e^{- i \lambda x } }{  
\pi \left( Ai( i b \lambda) Bi(a \Gamma+ i b \lambda)
- Bi( i b \lambda) Ai(a \Gamma+ i b \lambda) \right)}
\int_x^{+\infty} dy
 e^{ i \lambda y + \frac{\mu}{2  } y \Gamma - \frac{\mu^2}{12 } y^3 }
\end{eqnarray}
so that finally in terms of the scaling variables 
$\gamma$ (\ref{rescalegamma})
and $\tilde{x}$ (\ref{rescalx})
\begin{eqnarray}
V_{\Gamma}^R(x) && =  \frac{a}{\pi} e^{ \frac{1}{3} {\tilde x}^3 } 
\int_{-\infty}^{+\infty} \frac{d \lambda}{2 \pi } 
\frac{  e^{- i \lambda {\tilde x} } }{  
 \left( Ai( i  \lambda) Bi( \gamma+ i  \lambda)
- Bi( i  \lambda) Ai( \gamma+ i  \lambda) \right)}
\int_{\tilde x}^{+\infty}  d {\tilde y}
 e^{ i \lambda {\tilde y} + \gamma {\tilde y } - \frac{1}{3} {\tilde y}^3 }
\end{eqnarray}
Similarly, for the left half line $]-\infty,x]$, we have by symmetry 
\begin{eqnarray}
V_{\Gamma}^L(x) = V_{\Gamma}^R(-x)
\label{vsym}
\end{eqnarray}

In the limit $\gamma \to \infty$, we have 
using (\ref{asymptoktoy},\ref{asympAi}) and by
evaluating the integral over $y$ by the saddle-point method
\begin{eqnarray}
V_{\Gamma}^R(x) && \opsimeq_{\gamma \to \infty}
  \frac{a}{\pi} e^{ \frac{1}{3} {\tilde x}^3 } 
\int_{-\infty}^{+\infty} \frac{d \lambda}{2 \pi } 
\frac{  e^{- i \lambda {\tilde x} } }{  
 Ai( i  \lambda) Bi( \gamma+ i  \lambda)}
\int_{\tilde x}^{+\infty}  d {\tilde y}
 e^{ i \lambda {\tilde y} + \gamma {\tilde y } - \frac{1}{3} {\tilde y}^3 } \\
&&  \opsimeq_{\gamma \to \infty}
a  e^{ \frac{1}{3} {\tilde x}^3 } 
\int_{-\infty}^{+\infty} \frac{d \lambda}{2 \pi } 
\frac{ e^{- i \lambda {\tilde x} } }
 { Ai( i \lambda) } = e^{ \frac{1}{3} {\tilde x}^3 } g_{\infty}(x)
\end{eqnarray}
leading to 
\begin{eqnarray}
V_{\infty}(x) = g_{\infty}(-x) g_{\infty}(x)
\label{vinfty}
 \end{eqnarray}
as it should to recover the probability distribution
of the absolute minimum of the full landscape (\ref{distriminabs}).

In the other limit $\gamma \to 0$, we have using (\ref{wrAi})
\begin{eqnarray}
V_{\Gamma}^R(x) &&  \opsimeq_{\gamma \to 0} a e^{ \frac{1}{3} {\tilde x}^3 } 
\int_{-\infty}^{+\infty} \frac{d \lambda}{2 \pi } 
\frac{  e^{- i \lambda {\tilde x} } }{  
 \left(  \gamma+ \frac{\gamma^3}{3} i \lambda +... \right)}
\int_{\tilde x}^{+\infty}  d {\tilde y}
 e^{ i \lambda {\tilde y}  - \frac{1}{3} {\tilde y}^3 }
\left( 1+ \gamma {\tilde y } +...\right)  \\
&& \opsimeq_{\gamma \to 0} \frac{a}{\gamma} 
\left( 1 + \gamma {\tilde x } \right) + O(\gamma)
\label{vsmall}
\end{eqnarray}
and thus, for the points $x$ that
are in the region where $\tilde x$ remains finite, the probability (\ref{probavgamma}) is given by
\begin{eqnarray}
V_{\Gamma}(x)  \opsimeq_{\gamma \to 0} 
\frac{a^2}{ \gamma^2}  = \frac{1}{\Gamma^2}  
\label{resmindensitysinai}
\end{eqnarray}
i.e. it represents in this regime as it should
the uniform density $1/\Gamma^2$ of minima of the Sinai model \cite{us_long}.

Another quantity of interest is probability that $x$ is the 
nearest minimum 
$x^R_{min}$ from the boundary $x_R=+\infty$ in the renormalized
 landscape at scale $\Gamma$
\begin{eqnarray}
W^R_{\Gamma}(x)=V_{\Gamma}^L(x) V_{\Gamma}^R (x,+\infty,+\infty)
\label{firstmin}
\end{eqnarray}
where the constraint on the left half line
is the same as before (\ref{probavgamma},\ref{vsym})
and where the constraint on the right half-line reads (\ref{resggamma})
\begin{eqnarray}
V_{\Gamma}^R(x,\infty,\infty)  = 
\lim_{x_R \to \infty} \int_{\Gamma}^{+\infty} dF_R  
 E_{\Gamma}^-(0,x;F_R,x_R)
= e^{ \frac{\mu^2}{12 } x^3 } g_{\Gamma}(x)
\label{caseinfinite}
 \end{eqnarray}
or in terms of of the rescaled variables
using (\ref{resggamma},\ref{defrescalpsi})
\begin{eqnarray}
V_{\Gamma}^R(x,\infty,\infty)   
= a e^{ \frac{1}{3 } {\tilde x}^3 } 
\int_{-\infty} ^{+\infty} \frac{d\lambda}{2 \pi }
e^{- i \lambda {\tilde x}}
\frac{1}{ Ai(i  \lambda) } e^{- \int_{0}^{+\infty}
 df \tilde{\psi}_{\gamma} (f,\lambda)}
\label{caseinfiniterescal}
 \end{eqnarray}

Of course, in the regime $\gamma \to \infty$, the distribution (\ref{firstmin})
of $x^R_{min}$ is the same as the distribution of the absolute minimum (\ref{vinfty}). 
In the other limit $\gamma \to 0$, we expect that  
$x^R_{min}$ is much bigger than the scale $b$ (\ref{toyscale})
of the absolute minimum. In this regime, its position will not
be governed by the constraint on right landscape 
$V_{\Gamma}^R(x,\infty,\infty)$, which is easy to satisfy in this region,
but by the constraint on the left landscape, which behaves in this regime as
(\ref{vsmall})
\begin{eqnarray}
V_{\Gamma}^L(x) = V_{\Gamma}^R(-x)&& \opsimeq_{\gamma \to 0} \frac{a}{\gamma} 
\left( 1 - \gamma {\tilde x } \right) + O(\gamma)
\label{vsmall2}
\end{eqnarray}
We thus expect that ${\tilde x }^R_{min}$ scales as $1/\gamma$, i.e. 
\begin{eqnarray}
x^R_{min} \oppropto_{\gamma \to 0} \frac{b}{\gamma} 
\label{effectiveregion}
\end{eqnarray}
As a consequence, the total number $n$
of minima is the renormalized landscape at scale $\Gamma$
can be roughly estimated 
in this regime by dividing the effective region (\ref{effectiveregion})
by the typical distance $\Gamma^2$ (\ref{resmindensitysinai})
between two minima 
\begin{eqnarray}
n^{typ}(\gamma) \oppropto_{\gamma \to 0} \frac{1}{\gamma^3 }
\label{typicaltot} 
\end{eqnarray}
This estimation is coherent with the behavior found in (\ref{ptsmallequi}).
Of course, an exact calculation of the distribution of the number $n$
of minima of the renormalized landscape could in principle be done
from (\ref{decompoinfinite})
\begin{eqnarray}
Z_{free}=1=\sum_{n=1}^{+\infty} N^{(2n+2)}_{\Gamma} 
\end{eqnarray}
where $N^{2n+2}_{\Gamma}$ represents the total normalization
of the renormalized landscapes containing exactly $n$ local minima.
We have considered previously 
 $N^{(2)}_{\Gamma}$ (\ref{ptequi1}) corresponding to a single minimum, 
and $N^{(4)}_{\Gamma}$ (\ref{n4norma}) corresponding to two minima,
but unfortunately, the multiple integrals over the barriers
 $F$ and positions $x$
needed in (\ref{decompoinfinite}) become rapidly intractable
for arbitrary $n$ and arbitrary $\gamma$.

\subsection {Diffusion front}

As explained at the beginning of the Section \ref{dynamicstoy}, 
the long-time dynamics is determined by 
the properties of the renormalized landscape $\Gamma=T \ln t$,
and the results obtained will be valid in the 
 double limit $t \to \infty$ and $\mu \to 0$ 
in terms of the scaling variables $\gamma$ (\ref{rescalegamma})
and $\tilde{x}$ (\ref{rescalx}).
As in the Sinai model \cite{us_long}, the average over the disorder
of the diffusion front $\overline{ P(x,t \vert x_0,0) }$
is given by the probability that  $x$ is the minimum
 of a renormalized valley at scale 
  $\Gamma$ that contains the initial position $x_0$.

Let us first consider the case $x<x_0$.
The constraint for the left half-line $]-\infty,x]$ is simply given by $V_L(x)
=V_R(-x)$ (\ref{probavgamma},\ref{vsym}). The constraint for the landscape
on the right half line is now that $x$ should be the starting point of a renormalized bond
of arbitrary barrier $F \geq \Gamma$ and of length $l$ satisfying $x+l>x_0$.
We thus have 
\begin{eqnarray}
\theta(x_0-x) \overline{P(x,t \vert x_0,0) } = V_{\Gamma}^L(x) 
\left( V_{\Gamma}^R (x,+\infty,+\infty) 
+ \int_{\Gamma}^{+\infty} dF \int_{x_0-x}^{+\infty} dl \ V_R(x,F,l)
 \right)
\label{deffront}
\end{eqnarray}
where 
 $V_{\Gamma}^R$ represents the constraints
 on the landscape on the half line $[x,+\infty[$ :
the point $x$ should either be the first minimum 
on the right of the renormalized landscape (\ref{caseinfinite})
 or it should be the starting point of a finite ascending renormalized
 bond of barrier $F$ and length $l$, 
followed by an arbitrary descending bond, i.e. in terms of the 
Fokker-Planck block (\ref{changetotilde},\ref{globalshift})
\begin{eqnarray}
V_{\Gamma}^R(x,F,l) && =  B_{\Gamma}^-(0,x;F,x+l)
\int_{x+l}^{+\infty} dy B_{\Gamma}^+(F,x+l;F-\Gamma,y) \\
&& =  B_{\Gamma}^-(0,x;F,x+l)
\int_{x+l}^{+\infty} dy B_{\Gamma}^+(\Gamma,x+l;0,y)
\label{vxfldef}
 \end{eqnarray}

Using the explicit expressions (\ref{defhtoy},\ref{resblocbarrier},\ref{ktoy}),
we have 
\begin{eqnarray}
V_{\Gamma}^R(x,F,l) && = \frac{a^2}{\pi^2} e^{ \frac{\mu^2}{12 } x^3 + \frac{\mu}{2  }
 (F-\Gamma) (x+l)  }  \\
&& \int_{-\infty}^{+\infty} \frac{d \lambda_1}{2 \pi} 
\frac{ e^{ i \lambda_1 l + \int_0^{F-\Gamma} dv 
 a \left( \frac{  Ai'(a v + b i \lambda_1) Bi(a (v+\Gamma) + b i \lambda_1)
- Bi'(a v + b i \lambda_1) Ai(a (v+\Gamma) + b i \lambda_1)  }
{  Ai(a v + b i \lambda_1) Bi(a (v+\Gamma) + b i \lambda_1)
- Bi(a v + b i \lambda_1) Ai(a (v+\Gamma) + b i \lambda_1)  } \right) } }
{ \left( Ai(a (F-\Gamma) + b i \lambda_1) Bi(a F + b i \lambda_1)
- Bi(a (F-\Gamma) + b i \lambda_1) Ai(a F + b i \lambda_1) \right)  } \\
&& \int_{x+l}^{+\infty} dy  e^{ \ - \frac{\mu^2}{12 } y^3  } 
\int_{-\infty}^{+\infty} \frac{d \lambda_2}{2 \pi} 
\frac{ e^{ i \lambda_2 (y-x-l) } }
{ \left( Ai(  b p) Bi(a \Gamma + b i \lambda_2)
- Bi( b p) Ai(a \Gamma + b i \lambda_2) \right) }
\label{functionxfl}
 \end{eqnarray}
so that in terms of the scaling variables $\gamma$ (\ref{rescalegamma})
and $\tilde{x}$ (\ref{rescalx}), the integral needed in (\ref{deffront})
reads
\begin{eqnarray}
&&  \int_{\Gamma}^{+\infty} dF \int_{x_0-x}^{+\infty} dl \ V_R(x,F,l)
= \int_{\gamma}^{+\infty} \frac{df}{a} 
\int_{{\tilde x_0}-{\tilde x} }^{+\infty} b d{\tilde l} \
 V_{\Gamma}^R(x=b {\tilde x},F=\frac{f}{a} ,l=b {\tilde l})  \\
&& =  \frac{a}{\pi^2 } 
 e^{ \frac{1}{3 } {\tilde  x}^3 }
\int_{\gamma}^{+\infty} df 
\int_{{\tilde x_0}-{\tilde x} }^{+\infty}  d{\tilde l}
e^{  (f-\gamma) ({\tilde x}+ {\tilde l})  }  \\
&& \int_{-\infty}^{+\infty} \frac{d  \lambda_1}{2 \pi } 
\frac{ e^{ i \lambda_1 {\tilde l} + \int_0^{f-\gamma} dw 
  \left( \frac{  Ai'(w +  i \lambda_1) Bi(w+\gamma +  i \lambda_1)
- Bi'(w +  i \lambda_1) Ai(w+\gamma +  i \lambda_1)  }
{  Ai(w +  i \lambda_1) Bi(w+\gamma +  i \lambda_1)
- Bi(w +  i \lambda_1) Ai(w+\gamma +  i \lambda_1)  } \right) } }
{ \left( Ai(f-\gamma +  i \lambda_1) Bi(f +  i \lambda_1)
- Bi(f-\gamma +  i \lambda_1) Ai(f +  i \lambda_1) \right)  } \\
&& \int_{{\tilde x}+{\tilde l}}^{+\infty}  d {\tilde y}
  e^{ \ - \frac{1}{3 } {\tilde y}^3  } 
\int_{-\infty}^{+\infty} \frac{d \lambda_2}{2 \pi } 
\frac{ e^{ i \lambda_2 ({\tilde y}-{\tilde x}-{\tilde l}) } }
{ \left( Ai(  i \lambda_2) Bi( \gamma +  i \lambda_2)
- Bi( i \lambda_2) Ai( \gamma +  i \lambda_2) \right) }
\label{functionxflrescal}
 \end{eqnarray}

In the limit $\gamma \to \infty$, the term (\ref{caseinfinite}) dominates
and the diffusion front converges as it should towards the distribution
of the absolute minimum for $x$ and doesn't depend anymore on the initial
position $x_0$
\begin{eqnarray}
 \overline{P(x,t \vert x_0,0) } \opsimeq_{\gamma \to \infty}
g_{\infty}(-x) g_{\infty}(-x)
\end{eqnarray}

In the other limit $\gamma \to 0$, the term (\ref{caseinfinite})
is negligeable, and the typical length
associated to the dynamics is $\Gamma^2$ (\ref{resmindensitysinai}). 
The diffusion front is expected
to become in this regime a function of the scaling variable (\ref{rescalsinai}) 
\begin{eqnarray}
 (\Delta X)_{Sinai} = \frac{x_0-x} {\Gamma^2}
 = \frac{ {\tilde x_0} - {\tilde x} }{\gamma^2}  
\end{eqnarray}
The contribution of finite barriers
(\ref{functionxflrescal}) indeed becomes
with the appropriate changes of variables 
${\tilde l}= \gamma^2 l$, $\lambda_1= \frac{ \lambda}{\gamma^2}$,
$f=\gamma (1+\eta)$

\begin{eqnarray}
&&  \int_{\Gamma}^{+\infty} dF \int_{x_0-x}^{+\infty} dl \ V_R(x,F,l)
 \opsimeq_{\gamma \to 0}  \frac{a}{\pi  } 
 e^{ \frac{1}{3 } {\tilde  x}^3 }
\int_{0}^{+\infty}  d\eta 
\int_{\frac{{\tilde x_0}-{\tilde x}}{\gamma^2} }^{+\infty}  dl
e^{  \gamma \eta ({\tilde x}+ \gamma^2 l)  }  \\
&& \int_{-\infty}^{+\infty} \frac{d  \lambda}{2 \pi  } 
\frac{ e^{ i  \lambda l + \int_0^{ \eta} \gamma dv 
  \left( \frac{  Ai'(\gamma v +  i \frac{ \lambda}{\gamma^2}) 
Bi(\gamma v+\gamma +  i \frac{ \lambda}{\gamma^2})
- Bi'(\gamma v +  i \frac{ \lambda}{\gamma^2}) Ai(\gamma v+\gamma +  i \frac{ \lambda}{\gamma^2})  }
{  Ai(\gamma v +  i \frac{ \lambda}{\gamma^2}) 
Bi(\gamma v+\gamma +  i \frac{ \lambda}{\gamma^2})
- Bi(\gamma v +  i \frac{ \lambda}{\gamma^2}) Ai(\gamma v+\gamma +  i \frac{ \lambda}{\gamma^2})  } \right) } }
{ \left( Ai(\gamma \eta+  i \frac{ \lambda}{\gamma^2})
 Bi(\gamma (1+\eta) +  i \frac{ \lambda}{\gamma^2})
- Bi(\gamma \eta +  i \frac{ \lambda}{\gamma^2}) 
Ai(\gamma (1+\eta +  i \frac{ \lambda}{\gamma^2}) 
\right)  } 
  e^{ \ - \frac{1}{3 } ({\tilde x}+\gamma^2 l)^3  } \\
&& \opsimeq_{\gamma \to 0}  \frac{a}{ \gamma } 
\int_{0}^{+\infty}  d\eta 
\int_{\frac{{\tilde x_0}-{\tilde x}}{\gamma^2} }^{+\infty}  dl
 P^*(\eta,l)
  \label{sinaifront}
 \end{eqnarray}
where we have recognized the joint distribution of rescaled barrier $\eta$
and rescaled length $l$ of the Brownian landscape \cite{us_long}
\begin{eqnarray}
P^*(\eta,l)=  \int_{-\infty}^{+\infty} \frac{d  \lambda}{2 \pi  } 
\frac{ \left( i  \lambda \right)^{1/2} }{
\sinh \left( i  \lambda \right)^{1/2}} 
 e^{ i \lambda l -  \eta
 \left( i  \lambda \right)^{1/2}
\coth \left(  \left( i  \lambda \right)^{1/2} \right)   }  
\label{fpsinai}
 \end{eqnarray}
  
The diffusion front has thus for Laplace transform in this regime
\begin{eqnarray}
\int_x^{+\infty} dx_0 e^{-p (x_0-x) } \overline{P(x,t \vert x_0,0) }
\opsimeq_{\gamma \to 0} \int_0^{+\infty}  dy e^{-p \Gamma^2 y }
\int_{0}^{+\infty}  dl
\int_{0}^{+\infty}  d\eta 
 P^*(\eta,l) \frac{1- e^{-p \Gamma^2 l }}{p \Gamma^2}
= \frac{1}{\Gamma^2 p} \left(1-\cosh \Gamma \sqrt p  \right)
\end{eqnarray}
which is exactly the Laplace transform of the Kesten distribution for the Sinai model \cite{us_long}.

The diffusion front for $x_0<x$ can be obtained from the 
 symmetry $x \to -x$
\begin{eqnarray}
\theta(x-x_0) \overline{P(x,t \vert x_0,0) } && = \left( V_{\Gamma}^L(x,+\infty,+\infty) 
+ \int_{\Gamma}^{+\infty} dF \int_{x-x_0}^{+\infty} dl \ V_L(x,F,l)
 \right) V_{\Gamma}^R(x) \\
&& = \left( V_{\Gamma}^R(-x,+\infty,+\infty) 
+ \int_{\Gamma}^{+\infty} dF \int_{x-x_0}^{+\infty} dl \ V_R(-x,F,l)
 \right) V_{\Gamma}^R(x)
\label{deffront2}
\end{eqnarray}

\section{Results for stationary landscapes}

\label{statiolandscape}

In this section, we discuss the case of the stationary landscapes
defined by (\ref{defstatio}) which belong to the class
of solvable models with our method as discussed in Section (\ref{solvable}).

\subsection{Explicit expressions for the blocks}

We have seen in (\ref{transfostatio}) how the functional integrals for the process
$dU/dx =  - dW[U]/dU  + \eta(x)$ could be recast into a functional
integral of the Schrodinger-type {\it up to boundary terms}.
The Fokker-Planck blocks read (\ref{transfostatio})
\begin{eqnarray}
&& B^\pm_{\Gamma}( u_0, x_0 , u_1, x_1) =
e^{- \frac{1}{2} (W[u_1] - W[u_0])} 
\tilde{B}^\pm_{\Gamma}( u_0, x_0 , u_1, x_1)  \\
&& E^\pm_{\Gamma}( u_0, x_0 , u_1, x_1) =
e^{- \frac{1}{2} (W[u_1] - W[u_0])} 
\tilde{E}^\pm_{\Gamma}( u_0, x_0 , u_1, x_1)  
\label{defrealBstatio}
\end{eqnarray}
in terms of the Schrodinger blocks $\tilde{B}$ and $\tilde{E}$
given in Laplace transform by
\begin{eqnarray}
 \tilde{B}^-_{\Gamma} ( u_0, u,p) && = 
\tilde{B}^+_{\Gamma} ( u, u_0,p) = \frac{1}{K(u- \Gamma,u,p)} 
\exp( \int_{u_0 }^{u- \Gamma} dv \partial_1 \ln K(v,v+ \Gamma ,p) ) 
\\ \tilde{E}^-_{\Gamma} (u, u_b,p) && =  \tilde{E}^+_{\Gamma} (u_b, u,p) =
\exp(  \int_{u}^{u_b} dv  \partial_1 \ln K(v , v + \Gamma,p) )
\label{schrostatio}
\end{eqnarray}
The function $K$ (\ref{defK}) 
now corresponds to the problem 
with the Schrodinger potential $V[U]=- \frac{1}{2} W''[U] +
\frac{1}{4} W'[U]^2$.

For the special case $p=0$,
two independent solutions of the Schrodinger equation are then known
for an arbitrary potential $W[U]$
\begin{eqnarray}
&& \phi_1(u) = e^{-\frac{1}{2} W(u)}  \\
&& \phi_2(u) = e^{-\frac{1}{2} W(u)} \int_0^u e^{W(u')} du'
\label{phiforpeq0}
\end{eqnarray}
of wronskian $w=1$ and thus the function $K$ (\ref{defK}) simply reads :
\begin{eqnarray}
K(u,v,p=0) = e^{-\frac{1}{2} ( W(u) + W(v))} 
\int_{u}^{v} e^{W(u')} du'
\end{eqnarray}
The Schrodinger blocks of the measure read (\ref{schrostatio})
\begin{eqnarray}
&& {\tilde B}^-_{\Gamma}( u_0, u, p=0) = {\tilde B}^+_{\Gamma}( u, u_0, p=0) =
\frac{e^{ \frac{W(u_0)}{2} + \frac{W(u)}{2}}}{
\int_{u-\Gamma}^{u} e^{W(u')} du' }
e^{ - \int_{u_0}^{u-\Gamma} 
 \frac{dv}{\int_{v}^{v+\Gamma} e^{W(u')-W(v)} du'}  }\\
&& {\tilde E}^+_{\Gamma}( u_b, u, p=0)={\tilde E}^-_{\Gamma}( u, u_b, p=0) =
e^{ \frac{W(u)}{2} - \frac{W(u_b)}{2} - \int_{u}^{u_b} 
 \frac{dv}{\int_{v}^{v+\Gamma} e^{W(u')-W(v)} du'}}
\end{eqnarray}
so that finally the blocks for the full measure (\ref{defrealBstatio}) read
in terms of the arbitrary potential $W(u)$
\begin{eqnarray}
&& B^-_{\Gamma}( u_0, u, p=0) =
\frac{e^{ W(u_0)}  }{
\int_{u-\Gamma}^{u} e^{W(u')} du' }
e^{ - \int_{u_0}^{u-\Gamma} 
 \frac{dv}{\int_{v}^{v+\Gamma} e^{W(u')-W(v)} du'}  }  \\
&& B^+_{\Gamma}( u, u_0, p=0) =
\frac{e^{ W(u)}}{
\int_{u-\Gamma}^{u} e^{W(u')} du' }
e^{ - \int_{u_0}^{u-\Gamma} 
 \frac{dv}{\int_{v}^{v+\Gamma} e^{W(u')-W(v)} du'}  }\\
&& {\tilde E}^-_{\Gamma}( u, u_b, p=0) =
e^{ W(u)-W(u_b) - \int_{u}^{u_b} 
 \frac{dv}{\int_{v}^{v+\Gamma} e^{W(u')-W(v)} du'} }\\
&& {\tilde E}^+_{\Gamma}( u_b, u, p=0) =
e^{ - \int_{u}^{u_b} 
 \frac{dv}{\int_{v}^{v+\Gamma} e^{W(u')-W(v)} du'} }
\end{eqnarray}
These expressions for the Laplace variable
$p=0$ describe the properties of the extrema
in the energy $u$ of the landscape after integration
over the positions variables $x$.

\subsection{Normalization of the full measure}

The normalization for left and right reflecting boundaries,
corresponding to $u(x_L)=u_L$ and $u(x_R)=u_R$ with $u(x_L^-)=+\infty$ and $u(x_R^+)=+\infty$, is given by (\ref{fokker-planck})
\begin{eqnarray}
Z(u_L, x_L;u_R, x_R) =
\int_{u(x_L)=u_L}^{u(x_R)=u_R} DU
\exp(- \int_{x_L}^{x_R} dx ( \frac{1}{4} (\frac{dU}{dx} - F[U])^2 +
\frac{1}{2} F'[U] ))
\end{eqnarray}
and the decomposition into blocks at scale $\Gamma$ reads (\ref{decoupled})
\begin{eqnarray}
  Z(u_L,x_L,u_R,x_R) =  \sum_{n=1}^{+\infty} 
\int \prod_{i=1}^{2n-1} du_i dx_i 
&& E^+_{\Gamma} ( u_L, x_L ; u_1, x_1 )
B^-_{\Gamma} ( u_1, x_1 ;  u_2, x_2 )  \ldots \\
&& \ldots  B^+_{\Gamma} ( u_{2n-2}, x_{2n-2} ;  u_{2n-1}, x_{2n-1} )
E^-_{\Gamma} (  u_{2n-1}, x_{2n-1} ; u_R ,x_R ) \\
= e^{ - \frac{1}{2} (W[U(x_R)] -  W[U(x_L)]) } \sum_{n=1}^{+\infty} 
\int \prod_{i=1}^{2n-1} du_i dx_i
&& \tilde{E}^+_{\Gamma} ( u_L,  u_1, x_1-x_L )
\tilde{B}^-_{\Gamma} ( u_1,  u_2, x_2-x_1 )  \ldots \\
&& \tilde{B}^+_{\Gamma} ( u_{2n-2},   u_{2n-1}, x_{2n-1}- x_{2n-2} )
\tilde{E}^-_{\Gamma} (  u_{2n-1} , u_R ,x_R - x_{2n-1}) 
\label{normaz}
\end{eqnarray}

\subsection{Example : Brownian landscape with drift}

We rewrite the general expressions for the particular case $W(u)=f u$
corresponding to the Schrodinger potential
\begin{eqnarray}
V(u)=f^2/4=\delta^2
\label{potcte}
\end{eqnarray}
where we have introduced the notation $\delta=f/2$ to compare more straightforwardly
with the papers \cite{dsfrg}.
The functions $\phi_1(u,p)$ and $\phi_2(u,p)$ introduced in (\ref{defK}) 
and their wronskian $w(p)$ read
\begin{eqnarray}
&& \phi_1(u)=e^{- u \sqrt{p + \delta^2}} \\
&& \phi_2(u)=e^{u \sqrt{p + \delta^2}} \\
&& w(p)=2 \sqrt{p + \delta^2}
\end{eqnarray}
Thus the function $K$ introduced in (\ref{defK}) reads
\begin{eqnarray}
K(u,v,p) = \frac{1}{\sqrt{p + \delta^2}} \sinh((v-u) \sqrt{p + \delta^2})
\label{kbrown}
\end{eqnarray}
This leads to the expressions
\begin{eqnarray}
 B^-_{\Gamma}( u_1, u_2, p) && =
e^{- \delta (u_2-u_1) }
\frac{\sqrt{p + \delta^2}}{\sinh(\Gamma \sqrt{p + \delta^2})}
\exp( - (u_2-u_1-\Gamma) \sqrt{p + \delta^2} \coth(\Gamma \sqrt{p + \delta^2})) \\
&& = e^{- \Gamma \delta}
\frac{\sqrt{p + \delta^2}}{\sinh(\Gamma \sqrt{p + \delta^2})}
\exp( - (u_2-u_1-\Gamma) (\delta + \sqrt{p + \delta^2} \coth(\Gamma \sqrt{p + \delta^2}))) \\
 B^+_{\Gamma}( u_1, u_2, p) && =
e^{- \delta (u_2-u_1) }
\frac{\sqrt{p + \delta^2}}{\sinh(\Gamma \sqrt{p + \delta^2})}
\exp( - (u_1-u_2-\Gamma) \sqrt{p + \delta^2} \coth(\Gamma \sqrt{p + \delta^2})) \\
&& = e^{ \Gamma \delta}
\frac{\sqrt{p + \delta^2}}{\sinh(\Gamma \sqrt{p + \delta^2})}
\exp( - (u_1-u_2-\Gamma) (- \delta + \sqrt{p + \delta^2} \coth(\Gamma \sqrt{p + \delta^2}))) \\
 E^+_{\Gamma}( u_L, u) && = \exp( - 
(u_L-u) (- \delta + \sqrt{p + \delta^2} \coth(\Gamma \sqrt{p + \delta^2}))) \\
 E^-_{\Gamma}( u, u_R) && = \exp( - 
(u_R-u) (\delta + \sqrt{p + \delta^2} \coth(\Gamma \sqrt{p + \delta^2}))) 
\label{solupure}
\end{eqnarray}
in agreement with \cite{dsfrg}.
We thus recover simply and from a more general perspective the 
RSRG fixed-point for the pure Brownian landscape \cite{dsfrg,us_long}.

\subsection{Some further examples}

One can study cases where the potential $W(u)$ (\ref{defstatio})
is a simple power law $W(u)=r u^{\gamma}$. 
This corresponds to the Schrodinger potential (\ref{potstatio})
\begin{eqnarray}
V(U)= - \frac{r \gamma (\gamma-1)}{2} u^{\gamma-2} +\frac{r^2 \gamma^2}{4} u^{2(\gamma-1)}
\end{eqnarray}
which a priori contains two power-laws.
Apart from the case $\gamma=1$ corresponding to the biased Brownian motion
discussed above (\ref{potcte}), there are two other simple cases
 $\gamma=2$ and $\gamma=0$, which
we now briefly comment on.

\subsubsection{Landscape produced by a Brownian particle in a parabolic well}

The parabolic potential $W(u)=r u^2$
corresponds to the quadratic Schrodinger potential
\begin{eqnarray}
V(u)= -r+r^2 u^2  
\end{eqnarray}
The functions $\phi_1(u,p)$ and $\phi_2(u,p)$ introduced in (\ref{defK}) 
and their wronskian $w(p)$ read
\begin{eqnarray}
&& \phi_1(u,p)=e^{- \frac{r}{2} u^2} \sqrt{r} u 
 U(\frac{1}{2}+\frac{p}{4 r},\frac{3}{2},r u^2)\\
&& \phi_2(u,p)=e^{- \frac{r}{2} u^2}  u  
M(\frac{1}{2}+\frac{p}{4 r},\frac{3}{2},r u^2) \\
&& w(p)=  
\frac{\sqrt{\pi}}{\Gamma \left(\frac{1}{2}+\frac{p}{4 r}\right)}
\end{eqnarray}
where $U(a,b,z)$ and $M(a,b,z)$ are the confluent hypergeometric functions.
In particular for $p=0$, we have $U(\frac{1}{2} ,\frac{3}{2},z)=1/\sqrt{z}$ and (\ref{phiforpeq0}) yields
\begin{eqnarray}
&& \phi_1(u,p=0)=e^{- \frac{r}{2} u^2}  = e^{- \frac{1}{2} W(u) }  \\
&& \phi_2(u,p=0)=e^{- \frac{r}{2} u^2}  \int_0^u du' e^{r u'^2} \\
&& w(p=0)= 1 \\
&& K(u,v,p=0) = e^{ - \frac{r}{2} ( u^2 +v^2) }
\int_u^v du' e^{r u'^2} 
\end{eqnarray}

The above expressions allow to obtain the full joint distribution of
wide excursions (extrema) of the trajectory $U(x)$ of 
a Brownian particle in a quadratic well, but we will not
explore these results further here since this goes beyond the scope of
this paper. 

\subsubsection{Landscape produced by a Brownian particle in a logarithmic well} 

We now consider the logarithmic potential $W(u)= - r \ln  u $
for $u \in ]0,+\infty[$
corresponding to the Schrodinger potential
\begin{eqnarray}
V(u)= \left(- \frac{r}{2}+\frac{r^2}{4} \right) \frac{1}{u^2}
\end{eqnarray}

The functions $\phi_1(u,p)$ and $\phi_2(u,p)$ introduced in (\ref{defK}) 
and their wronskian $w(p)$ read
\begin{eqnarray}
&& \phi_1(u,p)= {\sqrt u} K_{\frac{1-r}{2}} ({\sqrt p} u) \\
&& \phi_2(u,p)= {\sqrt u} I_{\frac{1-r}{2}} ({\sqrt p} u)
 \\
&& w(p)= 1
\end{eqnarray}
where $K_{\nu}(z)$ and $I_{\nu}(z)$ are the Bessel functions. 
For the special case $p=0$,
(\ref{phiforpeq0}) yields for the function $K$ 
\begin{eqnarray}
K(u,v,p=0) = e^{-\frac{1}{2} ( W(u) + W(v))} 
\int_{u}^{v} e^{W(u')} du'= \frac{1}{1-r} \left[ u^{r/2} v^{1-r/2}- u^{1-r/2} v^{r/2} \right]
\end{eqnarray}

\subsection{An example with a long-ranged constraint : Positive Brownian
landscape }

As an example of long-ranged constraint, we consider
the process $U(x) = |V(x)|$
where $V(x)$ is a pure Brownian, i.e. $U(x)$ is constrained to remain positive
by a reflecting wall at $U=0$.
For fixed boundary conditions $(u_L,x_L)$ and $(u_R,x_R)$, the total normalization reads
\begin{eqnarray}
Z(u_L,x_L;u_R,x_R)= \frac{1}{\sqrt{4 \pi (x_R-x_L)}}
\left( e^{- \frac{(u_R-u_L)^2}{4 (x_R-x_L)} }
+  e^{- \frac{(u_R+u_L)^2}{4 (x_R-x_L)} }  \right) 
\end{eqnarray}
In Laplace transform with respect to the length $(x_R-x_L)$, this leads to
\begin{eqnarray}
Z(u_L,u_R;p)=\int_{x_L}^{+\infty} dx_R e^{-p (x_R-x_L)}
 Z(u_L,x_L;u_R,x_R)= \frac{1}{2 \sqrt{p}} \left(
e^{- \sqrt{p} \vert u_R-u_L \vert } + e^{- \sqrt{p} ( u_R-u_L ) } \right)
\label{mesinlaplace}
\end{eqnarray}

We again consider the convention $U(x_R^-)=U(x_L^+)=+\infty$
to define the renormalization procedure.
The decomposition of the renormalized landscape at scale $\Gamma$
can then be written as
\begin{eqnarray}
&&  Z(u_L,u_R;p)  = 
\sum_{n=1}^{+\infty}  
\int_{u_i>0}
E_{\Gamma}^+(u_L,u_1;p) B_{\Gamma}^- (u_1,u_2;p) ... 
E_{\Gamma}^- (u_{2n-1},u_R;p) \\
&& + E_{\Gamma}^+(u_L,0;p) G_{\Gamma} (0,0;p)  
E_{\Gamma}^- (0,u_R;p) \\
&& + E_{\Gamma}^+(u_L,0;p) G_{\Gamma} (0,0;p) 
\left( \sum_{n=1}^{+\infty}  
\int_{u_i>0}
B_{\Gamma}^-(0,u_1;p) B_{\Gamma}^+ (u_1,u_2;p) ... B_{\Gamma}^+ (u_{2n-1},0;p) \right) 
G_{\Gamma} (0,0;p) E_{\Gamma}^- (0,u_R;p)
+...
\label{positivebrown}
\end{eqnarray}
where the blocks ($B$,$E$) represents the `bulk' blocks
and are thus identical to the blocks for the pure Brownian motion
(\ref{solupure}) for $\delta=0$
\begin{eqnarray}
B_{\Gamma}^- (u_1,u_2;p)&& = B_{\Gamma}^+ (u_2,u_1;p) =
 \theta ((u_2-u_1)-\Gamma) \frac{ \sqrt{p} }{ \sinh(\Gamma \sqrt{p}) }
e^{- ((u_2-u_1)-\Gamma) \sqrt{p} \coth(\Gamma \sqrt{p}) } \\
E_{\Gamma}^- (u,u_b;p)&& = E_{\Gamma}^+ (u_b,u;p) =
 \theta (u_b-u) e^{- (u_b-u) \sqrt{p} \coth(\Gamma \sqrt{p}) }
\end{eqnarray}
and where the notation $G_{\Gamma} (u_1,u_2,p) $ represents 
the measure of the Brownian paths 
that remain in $[0,\Gamma]$ between $u_1$ and $u_2$.
Since $U(x) = |V(x)|$, this is equivalent to 
asking that the unconstrained Brownian $V(x')$ remains  
in the interval $]-\Gamma,\Gamma[$, and goes from $u_1$
to $(+u_2)$ or to $(-u_2)$. With the notations
of (\ref{basicpath},\ref{solubasicF}), its Laplace transform reads 
\begin{eqnarray}
G_{\Gamma} (u_1,u_2,p) && = 
F_{[u_a=-\Gamma, u_b=\Gamma]}(u_1,u_2,p) +
F_{[u_a=-\Gamma, u_b=\Gamma]}(u_1,-u_2,p)  \\
&& = \frac{K(-\Gamma,min(u_1,u_2)) K(max(u_1,u_2),\Gamma)}{K(-\Gamma,\Gamma)}
+ \frac{K(-\Gamma,-u_2) K(u_1,\Gamma)}{K(-\Gamma,\Gamma)}
\label{laplaceg00}
\end{eqnarray}
with the function $K$ given by  
$K(u,v,p) = \frac{1}{\sqrt{p }} \sinh((v-u) \sqrt{p })$ (\ref{kbrown}).
Here we only need the case $u_1=0=u_2$ 
\begin{eqnarray}
G_{\Gamma} (0,0,p)   
&& = \frac{\tanh(\Gamma \sqrt{p } )}{\sqrt{p }}
\end{eqnarray}
This expression may also be found by solving the  RSRG equation
\begin{eqnarray}
\partial_\Gamma G_{\Gamma}(0,0;p) = G_{\Gamma}(0,0;p) B_{\Gamma}(0,\Gamma,p) B_{\Gamma}(\Gamma,0,p) G_{\Gamma}(0,0;p) 
= G_{\Gamma}^2(0,0;p) \frac{p}{\sinh^2 \Gamma {\sqrt p}}
\end{eqnarray}
In the limit of $\Gamma \to \infty$, the decomposition 
(\ref{positivebrown}) becomes
\begin{eqnarray}
  Z(u_L,u_R;p) &&  = \int_0^{min(u_L,u_R)} du E_{\infty}^+(u_L,u;p)
E_{\infty}^- (u,u_R;p) + E_{\infty}^+(u_L,0;p) G_{\infty} (0,0;p)  
E_{\infty}^- (0,u_R;p) \\
&& = \int_0^{min(u_L,u_R)} du e^{- (u_L-u) \sqrt{p} } e^{- (u_R-u) \sqrt{p} }
+ e^{- u_L \sqrt{p} } \frac{1}{\sqrt{p }} e^{- u_R \sqrt{p} }
\end{eqnarray}
which coincides with (\ref{mesinlaplace}) as it should.

\newpage

\appendix

\section{Explicit construction of the measure for renormalized landscape} 

\label{calculsn2}

This Appendix is dedicated to the precise definitions
and properties of the joint probabilities of extrema
introduced in Section \ref{generalrg}.

\subsection{Initial landscape}

The joint probabilities defined in Section \ref{generalinitial}
are precisely defined as follows in terms of the initial
landscape measure (\ref{initialmeasure})
\begin{eqnarray}
&& N_{\text{init}}^{(2n)} \left(x_L,u_L ; u_1, x_1 ; \ldots u_{2n-1}, x_{2n-1} ; x_R, u_R \right)
= \int DU {\cal P}[U] \prod_{j=1}^{2n-1} \delta(U_{x_j}-u_j) \\
&& \prod_{i=1}^{x_1-1} \theta(U_{i} - U_{i+1})
\prod_{i=x_1}^{x_2-1} \theta(U_{i+1} - U_i)
\prod_{i=x_2}^{x_3-1} \theta(U_{i} - U_{i+1})
..
\prod_{i=x_{2n-1}}^{N-1} \theta(U_{i+1} - U_i)
\end{eqnarray}
The normalization (\ref{normainitiale}) can be checked by inserting
 the identity
\begin{eqnarray}
&& 1 = \prod_{i=1}^{N-1} (\theta(U_{i+1} - U_i) + \theta(U_{i} - U_{i+1})) \\
&& = \sum_{n=1}^{K} \sum_{1 \leq x_1 < x_2 < x_3 < .. < x_{2n-2} < x_{2n-1} \leq N}
\prod_{i=1}^{x_1-1} \theta(U_{i} - U_{i+1})
\prod_{i=x_1}^{x_2-1} \theta(U_{i+1} - U_i)
\prod_{i=x_2}^{x_3-1} \theta(U_{i} - U_{i+1})
..
\prod_{i=x_{2n-1}}^{N-1} \theta(U_{i+1} - U_i) \nonumber
\end{eqnarray}
(with $K=(N+1)/2$, $N$ is odd) into 
the initial measure (\ref{normameasure}).

\subsection{Renormalized landscape}

To define explicitly the measure of renormalized landscapes
defined in Section \ref{generalgamma}, it is convenient to
introduce the following notation for an ascending bond with $x_1<x_2$
\begin{eqnarray}
\Theta_\Gamma^-(x_1,x_2;U) = \prod_{i=x_1}^{x_2-1} \prod_{j=i+1}^{x_2} 
\theta(\Gamma - U_i + U_j) \theta(U_j - U_{x_1}) \theta(U_{x_2} - U_i) 
\end{eqnarray}
when $U_{x_2}-U_{x_1} \ge \Gamma$ and $0$ otherwise. It
expresses that $U_{x_1}$ is the minimum potential over the
segment $[x_1,x_2]$, $U_{x_2}$ is the maximum 
and all descending barriers in the segment with $i<j$
satisfy $U_i - U_j < \Gamma$.
Similarly, for a descending bond one defines
\begin{eqnarray}
\Theta_\Gamma^+(x_1,x_2;U) = \prod_{i=x_1}^{x_2-1} \prod_{j=i+1}^{x_2} 
\theta(\Gamma - U_j + U_i) \theta(U_{x_1} - U_j) \theta(U_i - U_{x_2}) 
\end{eqnarray}
when $U_{x_1}-U_{x_2} \ge \Gamma$ and $0$ otherwise. For $\Gamma=0$
we recover the extrema of the initial landscape
\begin{eqnarray}
&& \Theta_{\Gamma=0}^-(x_1,x_2;U) = \prod_{i=x_1}^{x_2-1} \theta(U_{i+1} - U_i) \\
&& \Theta_{\Gamma=0}^+(x_1,x_2;U) = \prod_{i=x_1}^{x_2-1} \theta(U_{i} - U_{i+1})
\end{eqnarray}
As a technicality, to treat the edges we must also introduce
\begin{eqnarray}
&& \tilde{\Theta}_\Gamma^+(x_1,x_2;U) = \prod_{i=x_1}^{x_2-1} \prod_{j=i+1}^{x_2} 
\theta(\Gamma - U_j + U_i) \theta(U_i - U_{x_2}) \\
&& 
\tilde{\Theta}_\Gamma^-(x_1,x_2;U) = \prod_{i=x_1}^{x_2-1} \prod_{j=i+1}^{x_2} 
\theta(\Gamma - U_i + U_j) \theta(U_j - U_{x_1})
\end{eqnarray}
with no other constraints, since outside the edge sites the potential 
is at $+ \infty$.
We may now express the probabilities of renormalized landscape 
in terms of the initial landscape as
\begin{eqnarray}
&& N_{\Gamma}  \left(x_L,u_L ; u_1, x_1 ; \ldots u_{2n-1}, x_{2n-1} ; x_R, u_R \right)
= \int DU {\cal P}[U] \prod_{j=1}^{2n-1} \delta(U_{x_j}-u_j) \\
&& \tilde{\Theta}_\Gamma^+(1,x_1;U)
\Theta_\Gamma^-(x_1,x_2;U) \Theta_\Gamma^+(x_2,x_3;U) ... 
\Theta_\Gamma^+(x_{2n-2},x_{2n-1};U)
\tilde{\Theta}_\Gamma^-(x_{2n-1},N;U)
\label{ngamma}
\end{eqnarray}
The normalization (\ref{normagamma}) can be checked by using
the following identity for any $\Gamma$
\begin{eqnarray}
&& 1 = \sum_{n=1}^{K} \sum_{1 \leq x_1 < x_2  ..  < x_{2n-1} \leq N}
\tilde{\Theta}_\Gamma^+(1,x_1;U)
\Theta_\Gamma^-(x_1,x_2;U) \Theta_\Gamma^+(x_2,x_3;U) ... 
\Theta_\Gamma^+(x_{2n-2},x_{2n-1};U)
\tilde{\Theta}_\Gamma^-(x_{2n-1},N;U)
\label{norm}
\end{eqnarray}
When $\Gamma$ is varied,
the following important property of renormalization 
\begin{eqnarray}
\partial_\Gamma \Theta_\Gamma^-(x_1,x_2;U) = 
\sum_{y_1=x_1+1}^{x_2-1} \sum_{y_2=y_1+1}^{x_2-1}
\Theta_\Gamma^-(x_1,y_1;U)
\Theta_\Gamma^+(y_1,y_2;U) \delta(U_{y_1} - U_{y_2} - \Gamma)
\Theta_\Gamma^-(y_2,x_2;U)
\label{rg}
\end{eqnarray}
leads to the renormalization of the full measure (\ref{rsrg}).

\subsection{Block product measures}

For the case of block product measure (\ref{decoupled}),
 the explicit expressions for the bond measures
and for the right edge read
\begin{eqnarray}
&& B^\pm_{\Gamma} ( u_2, x_2 ;  u_3, x_3 ) = 
\int \prod_{k=x_2}^{x_3} dU_k \delta(U_{x_2}-u_2) \delta(U_{x_3}-u_3)
\Theta_\Gamma^\pm(x_2,x_3;U) 
\exp( - \sum_{i=x_2}^{x_3-1} S_i[U_{i+1}, U_{i}] ) 
\label{block} \\
&& 
E^+_{\Gamma} ( x_R ;  u_1, x_1 ) = 
\int \prod_{k=1}^{x_1} dU_k \delta(U_{x_1}-u_1)
\tilde{\Theta}_\Gamma^+(1,x_1;U) 
\exp( - \sum_{i=1}^{x_1-1} S_i[U_{i+1}, U_{i}] ) \Phi_0(U_1)
\label{explidecoupled}
\end{eqnarray}
with a symmetric expression for the left edge.

\section{Renormalization procedure for periodic landscape}

In the Section \ref{generalrg}, we have described the renormalization procedure
for an interval $(x_L,x_R)$ with fixed boundary conditions.
In this Appendix, we briefly describe the changes that are necessary
for the case of a periodic landscape $U_1,...U_N$ with $U_{N+1}=U_1$.
The number of extrema is necessarily even in the initial landscape
as well as in the renormalized landscape.
 The renormalized landscape at scale 
$\Gamma$ can thus contain $2 n$ extrema at positions $1 \leq x_1 < x_2 <  ..< x_{2n} \leq N$
and $x_1$ can be either a minimum or a maximum. 
 As $\Gamma$ increases 
there are also samples such that all barriers in the systems are smaller than 
$\Gamma$, corresponding to $n=0$.

The normalization identity analogous to (\ref{norm})
which allows to treat the
periodic case reads
\begin{eqnarray}
&& 1 = G_0(\Gamma,U) + \sum_{n=1}^{K} 
\sum_{1 \leq x_1 < x_2 <  ..< x_{2n} \leq N} 
( \Theta_\Gamma^+(x_1,x_2;U) \Theta_\Gamma^-(x_2,x_3;U) ..
\Theta_\Gamma^+(x_{2n -1},x_{2 n};U) \Theta_\Gamma^-(x_{2 n},x_1;U) + \\
&& \Theta_\Gamma^-(x_1,x_2;U) \Theta_\Gamma^+(x_2,x_3;U) ..
\Theta_\Gamma^-(x_{2n -1},x_{2 n};U) \Theta_\Gamma^+(x_{2 n},x_1;U) )
\end{eqnarray}
where 
\begin{eqnarray}
G_0(\Gamma,U) = \prod_{i=0}^{N} \prod_{j=0}^{N} 
\theta(\Gamma - U_j + U_i) \theta(\Gamma - U_i + U_j)
\end{eqnarray}
is non zero if all barriers in the systems are smaller than 
$\Gamma$. 

Integrating over the measure $\int DU {\cal P}(U)$, normalized by $Z$, 
each term yields:
\begin{eqnarray}
Z = N_{\Gamma}^{(0)} +  \sum_{n=1}^{K} \int_{u_1,..u_{2 n}} \sum_{1 \leq x_1 < x_2 <  ..< x_{2n} \leq N} 
( N_{\Gamma}^{(2n,-)} \left(u_1, x_1 ; \ldots u_{2n}, x_{2n} \right) +
N_{\Gamma}^{(2n,+)} \left(u_1, x_1 ; \ldots u_{2n}, x_{2n} \right) )
\end{eqnarray}
where the symbols $\pm$ denote whether the bond $[x_{2n},x_1]$ is $\pm$. RSRG
recursion relations analogous to (\ref{rsrg})
may then be written, the only new
RSRG equation being the growing of $N_{\Gamma}^{(0)}$ 
upon the decimation of the samples with only two extrema left.

Block product measure, involving only $B^{\pm}_\Gamma$ can be also written
in an obvious way as in (\ref{decoupled}), and the above normalization identity can be written:
\begin{eqnarray}
Z = N_{\Gamma}^{(0)} +  \sum_{n=1}^{K} Tr (B^{+}_\Gamma B^{-}_\Gamma)^n + 
\sum_{n=1}^{K} Tr (B^{-}_\Gamma B^{+}_\Gamma)^n
\end{eqnarray}
each term, all integrations excluded, giving the desired $N_{\Gamma}^{(2n,\pm)}$.

Note that although the starting landscape here can be the same as the one
considered in Section \ref{generalrg} by imposing $U_1=U_N$ in
 (\ref{normameasure}), the
definition of the renormalized landscape here is different from the one
in Section \ref{generalrg} for fixed boundary conditions. 
Here one allows consideration of a bond $[x_{2n},x_1]$,
which can be decimated, 
whereas in the case with infinite left and right boundaries,
the decimation of the leftmost 
and rightmost bonds is not possible. 
In the periodic case, all the bonds can be decimated, 
resulting eventually in the
$G_0(\Gamma)$ factor.

\section{Derivation of the explicit solutions for a Schrodinger measure}

This Appendix contains the technical details of Section (\ref{explicitsolu}).

\subsection{Solution by a path-integral method}

\label{pathcomputation}

\subsubsection{Basic Path-integral}

\label{preli}

We first consider the following path integral  
\begin{eqnarray}
F_{[u_a,u_b]}(u_0, x_0, u , x ) = 
\int_{U_{x_0}=u_0}^{U_{x}=u} DU 
exp( - \int dx ( \frac{1}{4} (\frac{dU}{dx})^2 + V[U] )
\label{basicpath}
\end{eqnarray}
where the sum is over paths which
remain in the interval $]u_a,u_b[$ for $x_0<x<x_1$.
Seen as a function
of $(u,x)$, it satisfies the Schrodinger equation
\begin{eqnarray}
 \partial_x F = \partial^2_u F - V(u) F 
\end{eqnarray}
with the initial condition at $x=x_0$
\begin{eqnarray}
F(u,x_0)=\delta(u-u_0)  
\end{eqnarray}
and the absorbing boundary conditions at $u=u_a$ and $u=u_b$
\begin{eqnarray}
&& F(u_a,x) = 0 \\
&& F(u_b,x) = 0
\end{eqnarray}

The Laplace transform with respect to $(x-x_0)$
\begin{eqnarray}
F(u,p) = \int_{x_0}^{+\infty} dx e^{-p(x-x_0)} F(u,x)
\end{eqnarray}
 satisfies the system
\begin{eqnarray}
&& \partial^2_u F - V(u) F = p F - \delta(u-u_0) \\
&& F(u_a,p) = 0 \\
&& F(u_b,p) = 0
\label{systemF}
\end{eqnarray}

In terms of (\ref{defK}), we construct
two solutions of the equation (\ref{eqF})
which vanish respectively at $u_a$ and $u_b$ as:
\begin{eqnarray}
&& \Phi_-(u,p;u_a) = K(u_a,u,p) \\
&& \Phi_+(u,p;u_b) = K(u,u_b,p)
\end{eqnarray}
Their wronskian (also independent of $u$) read:
\begin{eqnarray}
 W(p ; u_a,u_b) && = \Phi_-'(u_0,p; u_a) \Phi_+(u_0,p ; u_b) 
- \Phi_-(u_0,p ; u_a)\Phi_+'(u_0,p ; u_b)  = K(u_a,u_b,p)
\end{eqnarray}

The solution of the above equations (\ref{systemF}) now read:
\begin{eqnarray}
\hat{F}_{[u_a,u_b]}(u,p|u_0) = \frac{1}{W(p ; u_a,u_b)} 
\Phi_-(u,p; u_a) \Phi_+(u_0,p ; u_b)
=  \frac{K(u_a,u,p) K(u_0,u_b,p)}{K(u_a,u_b,p)}
\qquad \text{if} ~~ u_a \leq u \leq u_0 \\
\hat{F}_{[u_a,u_b]}(u,p|u_0) = \frac{1}{W(p ; u_a,u_b)} 
\Phi_+(u,p ; u_b) \Phi_-(u_0,p ; u_a)
 = \frac{K(u_a,u_0,p) K(u,u_b,p)}{K(u_a,u_b,p)}
\qquad \text{if} ~~ u_0 \leq u \leq u_b 
\label{solubasicF}
\end{eqnarray}

The probability that the path has remained 
in the interval $]u_a,u_b[$ at time $x$
\begin{eqnarray}
S_{[u_a,u_b]}(x|u_0,x_0) =\int_{u_a}^{u_b} du
 F_{[u_a,u_b]}(u,x|u_0,x_0)
\end{eqnarray}
satisfies
\begin{eqnarray}
- \partial_x S_{[u_a,u_b]}(x|u_0,x_0) = 
r_{[u_a,u_b]}^{a}(x|u_0,x_0) + r_{[u_a,u_b]}^{b}(x|u_0,x_0)
\end{eqnarray}
where $r^{(u_a)}$ and $r^{(u_b)}$ represent the probability to be absorbed
between ''time'' $x$ and $x+dx$ at the boundary $u=u_a$ and
$u=u_b$ respectively. They read:
\begin{eqnarray}
&& r_{[u_a,u_b]}^{a}(x|u_0,x_0) =  \left[\partial_u F_{[u_a,u_b]}(u,x|u_0,x_0)
\right]|_{u=u_a} \\
&& r_{[u_a,u_b]}^{b}(x|u_0,x_0) = -  \left[\partial_u F_{[u_a,u_b]}(u,x|u_0,x_0)
\right]|_{u=u_b}
\end{eqnarray}
In Laplace transform with respect to $(x-x_0)$, they have the simple form
\begin{eqnarray}
\hat{r}^{a}_{[u_a,u_b]}(p;u_0 ) 
= \frac{\Phi_+(u_0,p ; u_b)}{\Phi_+(u_a,p ; u_b)} \\
\hat{r}^{b}_{[u_a,u_b]}(p;u_0 )
= \frac{\Phi_-(u_0,p ; u_a)}{\Phi_-(u_b,p ; u_a)}
\label{resrb}
\end{eqnarray}

We are now in position to evaluate the constrained path-integral
$B^-_{\Gamma} (u_1,x_1;u_2,x_2)$ (\ref{pathbm}).

\subsubsection{Expression of $B^-_{\Gamma} (u_1,x_1;u_2,x_2)$ for the case 
$ u_2=u_1 +\Gamma$}

\label{caseeqgamma}

We first consider the function $B^-_{\Gamma} (u_1,x_1;u_2,x_2)$
for an ascending bond with $u_2=u_1+\Gamma$.
Setting $u_1=u_a$, the distribution of $x_2$ is the 
distribution of the exit time at the boundary $u_b=u_a + \Gamma$ starting
from $u_0=u_a + \epsilon$ without being absorbed by the
boundary at $u_a$. We  thus have to consider the exit probability $r^{b}_{[u_a,u_b]}(x|u_0,x_0)$
computed in (\ref{resrb}), for the particular case $u_0=u_a+\epsilon$, $u_b=u_a+\Gamma$
\begin{eqnarray}
\lim_{\epsilon \to 0} \left( \frac{1}{\epsilon} r^{b}_{[u_a,u_a+\Gamma]}(x|u_a+\epsilon,x_0) \right) = 
B^{-}_{[u_a,u_a+\Gamma]}(x-x_0 )
\end{eqnarray}
where the function $B^{-}_{[u_a,u_a+\Gamma]}(x-x_0 )$ is defined by its
Laplace transform
\begin{eqnarray} 
\hat B^{-}_{[u_a,u_a+\Gamma]}(p) &&
= \partial_{u_0} \hat{r}^{b}_{[u_a,u_b]}(p;u_0 ) \vert_{u_0=u_a,u_b=u_0+\Gamma} 
  = \frac{1}{K(u_a,u_a+\Gamma,p)}
\label{bmgamma}
\end{eqnarray}

\subsubsection{Expression of $B^-_{\Gamma} (u_1,x_1;u_2,x_2)$ for the case $u_2>u_1 + \Gamma$}

We now consider the function $B^-_{\Gamma} (u_1,x_1;u_2,x_2)$
for an ascending bond with $u_2 > u_1+\Gamma$.
To take into account the constraint of no returns of more than $\Gamma$,
we write the recursion equation
\begin{eqnarray}
B^-_{\Gamma} ( u_0, x_0 ; u + \Delta u, x ) = 
\int_{x_0}^{x} dx'
B^-_{\Gamma} ( u_0, x_0 ; u , x' ) r^b_{[u_a=u-\Gamma,u_b=u+\Delta u]}(x|u,x')
\label{eqdiffBm}
\end{eqnarray}
i.e. we obtain in Laplace with respect to $(x-x_0)$ the differential equation
\begin{eqnarray}
\partial_u \hat B^-_{\Gamma} ( u_0,  ; u  ; p ) = 
\hat B^-_{\Gamma} ( u_0,  ; u ; p ) \partial_{u_b} r^b_{[u_a,u_b]}(p,u)
\vert_{u_a=u-\Gamma,u_b=u}
= \hat B^-_{\Gamma} ( u_0,  ; u ; p ) Z^{-}(p,u-\Gamma,u)
\end{eqnarray}
with 
\begin{eqnarray}
Z^{-}(p,u_1,u_2) = - \frac{\partial_{u_2} K(u_1,u_2,p)}{K(u_1,u_2,p)}
\end{eqnarray}
Thus using the initial condition at $u_1=u_0+\Gamma$ (\ref{bmgamma}), we get

\begin{eqnarray}
 B^-_{\Gamma} ( u_0, u,p) && = B^-_{\Gamma} ( u_0, u_0 + \Gamma,p)
\exp(  \int_{u_0 + \Gamma}^{u}  du' Z^-(p,u' - \Gamma,u' ) )
\end{eqnarray}
leading to (\ref{resbmpath}) given in the text.

\subsubsection{Edge bonds}

We now consider the function $E^{-}_{\Gamma}(u,x;u_R,x_R)$
for the ascending edge bond near the boundary condition $U(x=x_R^+)=+\infty$.
Given $u_R$, the only constraint is that
$u \leq u_R$ is the deepest minimum with no descending return of more than $\Gamma$
in between. If $u=u_R$, we simply have $x=x_R$, whereas for $u < u_R$
we may again write an equation similar to (\ref{eqdiffBm})
and we finally obtain in Laplace variable with respect to $(x_R-x)$
the result (\ref{resempath}) given in the text.

\subsection{Solution from the RSRG equations
 : Factorized Ansatz and Liouville equation}

\label{appliouville}

The factorized Ansatz (\ref{formdec}) for
$B^{-}_{\Gamma}$ leads to the following RG equation 
\begin{eqnarray}
\partial_{\Gamma} \ln B^{-}_{\Gamma} ( u ;  u'' ; p) 
= \int_{u}^{u''-\Gamma} du'
(B^{-}_{\Gamma} ( u' ; u'+ \Gamma ; p ))^2
\label{b2}
\end{eqnarray}
i.e. from the knowledge of $B^{-}_{\Gamma} ( u ; u + \Gamma ; p)$ 
for all $(\Gamma,u)$ one can reconstruct $B^{-}_{\Gamma} ( u ;  u'' ; p)$
for all $(\Gamma,u,u'')$. 
This RG equation can be integrated if we introduce an auxiliary function
$\Phi(u,v)$ defined by (\ref{carre})
since it yields
\begin{eqnarray}
&& \int_{u}^{u''-\Gamma} du' \partial_1 \partial_2 \Phi(u' , u' + \Gamma)
= \int_{u + \Gamma}^{u''} dv \partial_1 \partial_2 \Phi(v - \Gamma, v) \\
&& = - \partial_\Gamma ( \int_{u + \Gamma}^{u''} du' \partial_2 \Phi(u' - \Gamma, u') )
- \partial_\Gamma \Phi(u , u + \Gamma)
\end{eqnarray}

and thus, integrating (\ref{b2}) over $\Gamma$ one gets
\begin{eqnarray}
 B^{-}_{\Gamma} ( p ; u ;  u'') &&  = \exp(- \Phi(u , u + \Gamma))
\exp( - \int_{u + \Gamma}^{u''} du' \partial_2 \Phi(u' - \Gamma, u') ) 
\\
&&  = \exp(- \Phi(u''-\Gamma , u''))
\exp( - \int_{u'' - \Gamma}^{u} du' \partial_1 \Phi(u', u' + \Gamma) )
\label{solu}
\end{eqnarray}
which is indeed of the form (\ref{formdec}) with
\begin{eqnarray}
&& A^{L}_\Gamma(u) = \exp( - \int^{u} du' \partial_1 \Phi(u', u' + \Gamma) ) \\
&& A^{R}_\Gamma(u'') = \exp(- \Phi(u''-\Gamma , u'')) 
\frac{1}{A^{L}_\Gamma(u''-\Gamma)}
\end{eqnarray}
The consistency of (\ref{carre}) now requires 
the famous Liouville equation (\ref{liouville}) for the function $\Phi(u_1, u_2)$.

As is well known its most general
local solution is of the form (Liouville 1850) 
\begin{eqnarray}
\Phi(u_1, u_2) = \ln (\frac{f_L(u_1)- f_R(u_2)}{(f_L'(u_1) f_R'(u_2))^{1/2}} )
\end{eqnarray}
where $f_L(u)$ and $f_R(u)$ are two arbitrary analytic functions.
The alternative form (\ref{parawronsk}) can be obtained by noting
that the Liouville equation (\ref{liouville}) 
implies
\begin{eqnarray}
&& \partial_1 (e^{-\Phi} \partial_2^2 e^{\Phi}) = 0 \\
&& \partial_2 (e^{-\Phi} \partial_1^2 e^{\Phi}) = 0
\end{eqnarray}
and thus there must exist two functions $V_L(u)$ and $V_R(u)$ such that
$e^{\Phi}$ satisfies the two corresponding Schrodinger equations
\cite{gervais_neveu}. 
The connexion between the two representations is
made by noting that $f_L(u)=\psi_L^{(1)}(u)/\psi_L^{(2)}(u)$.
$f_R(u)=\psi_R^{(1)}(u)/\psi_R^{(2)}(u)$ and 
\begin{eqnarray}
&& \psi_L^{(1)}(u) = (- w_L)^{1/2} \frac{f_L(u)}{\sqrt{f_L'(u)}} \\
&& \psi_L^{(2)}(u) = (- w_L)^{1/2} \frac{1}{\sqrt{f_L'(u)}}
\end{eqnarray}
and similarly for $\psi_R$ and $f_R$.
One also has $V_L(u) = - \frac{1}{2} ( \frac{f_L'''(u)}{f_L'(u)} -
\frac{3}{2} (\frac{f_L''(u)}{f'_L(u)})^2)$ and in general a
second solution of the Schrodinger equation can be obtained from
a first one $\psi_L^{(2)}(u)$ as $\psi_L^{(1)}(u) = \psi_L^{(2)}(u)
\int^u (\psi_L^{(2)}(u))^{-2}$. Finally one finds that one can rewrite
(\ref{solu}) as
\begin{eqnarray}
 B^{-}_{\Gamma} ( p ; u ;  u'') &&= 
\frac{(f'_L(u) f'_R(u''))^{1/2}}{f_L(u) - f_R(u + \Gamma)}
\exp( \int_{f_R(u + \Gamma)}^{f_R(u'')} \frac{df_R}{f_L(u(f_R) - \Gamma) - f_R} ) \\
&& = \frac{(f'_L(u) f'_R(u''))^{1/2}}{f_L(u''-\Gamma) - f_R(u'')}
\exp( \int_{f_L(u)}^{f_L(u''-\Gamma)} \frac{df_L}{f_L - f_R(v(B_L) + \Gamma)} )
\label{soluB}
\end{eqnarray}
where $u(B_R)$ is the inverse function of $f_R(u)$, i.e.
$f_R(u(y))=y$ and $v(f_L)$ the inverse function of $f_L(u)$.

Finally, assuming that $B^{-}_{\Gamma} ( u ; u'' ; p) = A^{(1)}_\Gamma(u) A^{(2)}_\Gamma(u'')$
together with delta function initial conditions,
yields for the edge blocks 
\begin{eqnarray}
&& E^{+}_{\Gamma} (u_L, u,p) = \frac{A^{L}_\Gamma(u)}{A^{L}_\Gamma(u_L)} \\
&& E^{-}_{\Gamma} (u, u_R,p) = \frac{A^{L}_\Gamma(u)}{A^{L}_\Gamma(u_R)}
\end{eqnarray}

\section{Configurations presenting two nearly degenerate minima}

\label{twodege}

\subsection{Distribution of the absolute minimum for the toy model on an arbitrary interval}

The absolute minimum of the toy model on a arbitrary interval $[x_L,x_R]$
with fixed boundary conditions $(u_L,u_R)$ is given by
the $\Gamma \to \infty$ limit of the renormalized landscape 
containing only one valley left (\ref{decompofixed})
\begin{eqnarray}
&&  Z(u_L,x_L,u_R,x_R)  =  \int_{-\infty}^{min(u_L,u_R)} du 
\int_{x_L}^{x_R} dx N^{(2)}_{\infty} (x_L,u_L-u,x,u_R-u,x_R) 
\end{eqnarray}
where
\begin{eqnarray}
&& N^{(2)}_{\infty} (x_L,u_L-u,x,u_R-u,x_R)  = \nonumber  \\
&&  e^{ \frac{\mu^2}{12} x_L^3 -\frac{\mu}{2} x_L (u_L-u) } 
\tilde{E}^+_{\infty}( u_L-u, 0, x-x_L) 
 e^{ + \frac{\mu}{2} x_R (u_R-u) - \frac{\mu^2}{12} x_R^3 }
\tilde{E}^-_{\infty}( 0,  u_R-u, x_R-x) 
\label{defMtoy}
\end{eqnarray}
describes the statistical properties of
the position $x$ and the energy $u$ 
of the absolute minimum of the interval
given the boundary conditions $(u_L,x_L)$ and $(u_R,x_R)$.
To obtain the edge blocks $\tilde{E}$ (\ref{resblocbarrier})
in the limit $\Gamma \to \infty$, we need the function
$K(u_1,u_2)$ in the regime $u_2 = u_1+\Gamma \to \infty$, with $u_1$ fixed.
Using the asymptotic expressions of Airy functions (\ref{asympAi}),
we thus have
\begin{eqnarray}
 K(u_1,u_2) \opsimeq_{u_2 \to \infty} 
  \frac{\pi}{a} Ai(a u_1 + b p) Bi(a u_2 + b p)
\label{asympKtoy}
\end{eqnarray}
and the edge blocks at $\Gamma=\infty$
are simply given by
\begin{eqnarray}
 \tilde{E}^+_{\infty} (F,0,p) = \tilde{E}^-_{\infty} (0,F,p)
= \frac{Ai(a F + b p)}{Ai(b p)} 
\label{Etoylimit}
\end{eqnarray}
The function $N^{(2)}_{\infty}$ (\ref{defMtoy})
 describing the distribution of the minimum
thus reads
\begin{eqnarray}
&&   N^{(2)}_{\infty} (x_L,u_L-u,x,u_R-u,x_R)  = 
  e^{ \frac{\mu^2}{12} (x_L^3-x_R^3) 
 + \frac{\mu}{2} x_R (u_R-u) -\frac{\mu}{2} x_L (u_L-u) } \nonumber \\
&& \int_{-\infty}^{+\infty} \frac{d\lambda_1}{2 \pi} 
e^{ i \lambda_1 (x-x_L)} \frac{Ai(a (u_L-u) + b i \lambda_1)}{Ai(b i \lambda_1)}
\int_{-\infty}^{+\infty} \frac{d\lambda_2}{2 \pi}
e^{ i \lambda_2 (x_R-x)} 
\frac{Ai(a (u_R-u) + b i \lambda_2)}{ Ai(b i \lambda_2)}
\label{resMtoy}
\end{eqnarray}

\subsection{Probability to have two nearly degenerate minima}

To study the measure of configurations with two degenerate minima,
we need to consider (\ref{resMtoy}) for the special case
where the absolute minimum $u$ coincides with the left
boundary energy $u_L$
\begin{eqnarray}
&&   N^{(2)}_{\infty} (x_L,0,x,u_R-u_L,x_R)  = 
  e^{ \frac{\mu^2}{12} (x_L^3-x_R^3) 
 + \frac{\mu}{2} x_R (u_R-u_L) } \delta(x-x_L) 
\int_{-\infty}^{+\infty} \frac{d\lambda_2}{2 \pi}
e^{ i \lambda_2 (x_R-x)} 
\frac{Ai(a (u_R-u_L) + b i \lambda_2)}{ Ai(b i \lambda_2)}
\label{auxi2deg}
\end{eqnarray}

For the case $u_R=u_L+\epsilon$ with $\epsilon \to 0$, 
this yields for $x_L < x_R$
\begin{eqnarray}
  N^{(2)}_{\infty} (x_L,0,x,\epsilon,x_R)  = 
&& \delta(x-x_L)  e^{ \frac{\mu^2}{12} (x_L^3-x_R^3) }
\left(1 + \frac{\mu}{2} x_R \epsilon +O(\epsilon^2) \right)   \nonumber \\
&& \left( \delta(x_R-x)+ 
a \epsilon \int_{-\infty}^{+\infty} \frac{d\lambda_2}{2 \pi}
e^{ i \lambda_2 (x_R-x)} 
\frac{ Ai'(  b i \lambda_2)}{ Ai(b i \lambda_2)} +O(\epsilon^2)\right) \\
&& = a \epsilon \delta(x-x_L)  e^{ \frac{\mu^2}{12} (x_L^3-x_R^3) }
\int_{-\infty}^{+\infty} \frac{d\lambda_2}{2 \pi}
e^{ i \lambda_2 (x_R-x_L)} 
\frac{ Ai'(  b i \lambda_2)}{ Ai(b i \lambda_2)}
\label{auxiexpansion}
\end{eqnarray}

 For the case $x_R \to \infty$, after integration over the barrier
$(u_R-u_L)$, the formula (\ref{auxi2deg}) becomes using (\ref{Etoylimit},\ref{resggamma},\ref{defg}) 
\begin{eqnarray}
&&   N^{(2)}_{\infty} (x_L,0,x,+\infty)  \equiv
\lim_{x_R \to +\infty} \int_{u_L}^{+\infty} du_R  N^{(2)}_{\infty} (x_L,0,x,u_R-u_L,x_R) =   e^{ \frac{\mu^2}{12} x_L^3  } \delta(x-x_L) 
g_{\infty}(x_L)
\label{minL}
\end{eqnarray}
and similarly
\begin{eqnarray}
&&   N^{(2)}_{\infty} (+\infty,x,0,x_R) 
=  e^{ - \frac{\mu^2}{12} x_R^3  } \delta(x-x_R) 
g_{\infty}(-x_R)
\label{minR}
\end{eqnarray}

Putting together the three results (\ref{auxiexpansion},\ref{minL},
\ref{minR}), we get the formula (\ref{restwomin}) for the configurations
with two nearly degenerate minima.

\section{Study of two valleys configurations in the toy model}

\label{2valleys}

This Appendix is dedicated to the study of the two valleys configurations
described by (\ref{fulln4}). Since the barriers $F_1$ and $F_3$ 
are larger than $\Gamma$ satisfying
$a \Gamma \gg 1$, we may use the asymptotic expression (\ref{asympAi})
for the $B_i$ to get

\begin{eqnarray}
&& N^{(4)}_{\Gamma}(x_1,F_1,x_2,F_2,x_3) 
 \opsimeq_{\Gamma \to \infty} 
\frac{ e^{ \frac{\mu}{2} x_{2} (F_1-F_2) } }{\pi^2}
 \int_{-\infty}^{+\infty} \frac{d\lambda_1}{2 \pi} 
 \int_{-\infty}^{+\infty} \frac{d\lambda_2}{2 \pi} 
\int_{-\infty}^{+\infty} \frac{d\lambda_3}{2 \pi} 
\int_{-\infty}^{+\infty} \frac{d\lambda_4}{2 \pi} \\
&& \frac{e^{  i (\lambda_1+\lambda_2) x_1 +  i (\lambda_3-\lambda_2) x_2 
-i (\lambda_2+\lambda_3)x_3 }}{ Ai(b i \lambda_1) Ai(b \lambda_2)
 Ai(b \lambda_3)  Ai(b i \lambda_4)}
(a F_1 + b i \lambda_2')^{1/4} (a F_3 + b i \lambda_3')^{1/4} 
e^{-\frac{2}{3} (a F_1 + b i \lambda_2')^{3/2}
-\frac{2}{3} (a F_3 + b i \lambda_3')^{3/2}} 
\label{fulln4bis}
\end{eqnarray}

At large $\Gamma$, the scaling region for the barrier is thus
 \begin{eqnarray}
F=\Gamma + \frac{\eta} { a \sqrt{ a \Gamma}}
\label{barrierscaling}
\end{eqnarray}
with $\eta \ge 0$ of order 1. The barriers are thus concentrated in the vicinity
of $\Gamma$.
Integration over the two barriers $F_1,F_2$ yields at leading order after
the rescaling (\ref{barrierscaling}) 
\begin{eqnarray}
&&  \int_{\Gamma}^{+\infty} dF_1
\int_{\Gamma}^{+\infty} dF_2 N^{(4)}_{\Gamma}(x_1,F_1,x_2,F_2,x_3) 
\opsimeq_{\Gamma \to \infty} \\
&&  \frac{ a^2 (a \Gamma)^{-1/2} }{\pi} 
e^{-\frac{4}{3} (a \Gamma)^{3/2} }
\frac{1}{1+\frac{(a^2  x_2)^2}{a \Gamma} }
g_{\infty}(-x_1) 
 g_{\infty}(x_1-x_2 + b (a \Gamma)^{1/2} ) 
g_{\infty}(x_2-x_3 + b (a \Gamma)^{1/2} ) g_{\infty}(x_3)
\label{reshapprox}
\end{eqnarray}
in terms of (\ref{defg}).

We now integrate over the maximum $x_2$. The constraints of integration
$x_1<x_2<x_3$ may be forgotten since the domains $x_2<x_1$ and $x_2>x_3$
are highly suppressed with the functions
 $g_{\infty}(x_1-x_2 + b (a \Gamma)^{1/2} )$ and 
$g_{\infty}(x_2-x_3 + b (a \Gamma)^{1/2} )$. 
The measure for the two minima thus read
\begin{eqnarray}
M^{(4)}_{\Gamma}(x_1,x_3) && \equiv \int_{x_1}^{x_3} dx_2  \int_{\Gamma}^{+\infty} dF_1
\int_{\Gamma}^{+\infty} dF_2 N^{(4)}_{\Gamma}(x_1,F_1,x_2,F_2,x_3)  \opsimeq_{\Gamma \to \infty} 
\frac{  g_{\infty}(-x_1) 
g_{\infty}(x_3) 
 }{a^2}  e^{-\frac{4}{3} (a \Gamma)^{3/2} } \\
 &&
\int_{-\infty}^{+\infty} \frac{d\lambda_1}{2 \pi} 
\frac{e^{ - i \lambda_1 (x_1 + b (a \Gamma)^{1/2}) }}
{ Ai(b i \lambda_1)}
\int_{-\infty}^{+\infty} \frac{d\lambda_2}{2 \pi} 
\frac{e^{ - i \lambda_2 (-x_3 + b (a \Gamma)^{1/2}) }}
{ Ai(b i \lambda_2)}
e^{- b (a \Gamma)^{1/2}  \vert \lambda_1 - \lambda_2\vert}
\end{eqnarray}
which yields (\ref{restwomin}) for $a \Gamma \gg 1$.


\section{Useful properties of Airy functions}

\label{appairy}

The Airy functions $Ai$ and $Bi$ are 
 two independent solutions of the differential equation
\begin{eqnarray}
\phi''(z)=z \phi(z)
\label{eqAi}
\end{eqnarray}
of Wronskian
\begin{eqnarray}
Ai Bi' - Bi Ai' = \frac{1}{\pi}
\label{wrAi}
\end{eqnarray}
An immediate consequence is
\begin{eqnarray}
\frac{d}{dx} \left( \frac{Bi(x)}{Ai(x)} \right) =
\frac{1}{ \pi Ai(x)^{2} } 
\label{integai1}
\end{eqnarray}

The asymptotic expressions of the Airy functions at large argument read 
\begin{eqnarray}
&& Ai(z) \opsimeq_{ Re(z) \to +\infty} 
\frac{1}{2 \sqrt \pi} z^{- \frac{1}{4}} e^{-\frac{2}{3} z^{3/2}} \\
&& Bi(z) \opsimeq_{ Re(z) \to +\infty}
 \frac{1}{\sqrt{\pi}} z^{-1/4} e^{\frac{2}{3} z^{3/2}}
\label{asympAi}
\end{eqnarray}

Another useful property is
\begin{eqnarray}
\lim_{x \to \pm \infty} \frac{Bi(i x)}{Ai(i x)}= \pm i
\label{asympAiimag}
\end{eqnarray}


\end{document}